\documentclass[a4paper,11pt]{article}
\pdfoutput=1 
\usepackage{amssymb}
\usepackage{bbm}
\usepackage{siunitx}
\usepackage{amsmath}
\usepackage{braket}
\usepackage{mathtools}
\usepackage[usenames,dvipsnames,svgnames,table]{xcolor}
\usepackage [utf8] {inputenc}
\usepackage{color}
\usepackage{subcaption}
\usepackage{lmodern}
\usepackage{footnote}
\usepackage{slashed}
 \usepackage{mathrsfs}
 \usepackage[pdftex] {graphicx}
\usepackage{multirow}
\usepackage{jheppub} 
\usepackage{here}
\usepackage{hyperref}
\usepackage[justification=centering, singlelinecheck=false]{caption}
\usepackage[list=true, labelfont=bf, labelformat=brace, position=top]{subcaption}

\newcommand{\eq}{\begin{eqnarray}}

\newcommand{\en}{\end{eqnarray}}

\title{Relativistic-invariant formulation of the NREFT three-particle quantization condition}
\author[1]{Fabian M\"uller,}
\affiliation[1]{Helmholtz-Institut f\"ur Strahlen- und Kernphysik (Theorie) and\\ Bethe Center for Theoretical Physics, Universit\"at Bonn, 53115 Bonn, Germany}

\emailAdd{f.mueller@hiskp.uni-bonn.de}

\author[2]{Jin-Yi Pang,}
\affiliation[2]{College of Science, University of Shanghai for Science and Technology, Shanghai 200093, China}

\emailAdd{jypang@usst.edu.cn}

\author[1,3]{Akaki Rusetsky,}
\affiliation[3]{Tbilisi State  University,  0186 Tbilisi, Georgia}

\emailAdd{rusetsky@hiskp.uni-bonn.de}

\author[4]{and Jia-Jun Wu}
\affiliation[4]{School of Physical Sciences, University of Chinese Academy of Sciences, Beijing 100049, China}

\emailAdd{wujiajun@ucas.ac.cn}

\abstract{
  \begin{sloppypar}
    \noindent
    A three-particle quantization condition on the lattice is written down
    in a 
  manifestly
  relativistic-invariant form by using a generalization of the
  non-relativistic effective field
  theory (NREFT) approach. Inclusion of the higher partial waves
  is explicitly addressed.  A partial diagonalization of the quantization
  condition into the various irreducible representations of the (little groups of the) octahedral group has been carried out both in the center-of-mass frame and in moving frames. Furthermore, producing synthetic data in a toy model,
  the relativistic invariance is explicitly demonstrated for the three-body
  bound state spectrum.
  \end{sloppypar}
}

\allowdisplaybreaks
\begin{document}
\maketitle
\flushbottom

\section{Introduction}

Recent years have witnessed a rapid growth of interest to the three-body
problem on the lattice. This interest dates back to 2012, when it was shown,
for the first time, that the three-body spectrum in a finite volume is determined solely by the three-body $S$-matrix elements~\cite{Polejaeva:2012ut}. In the next years, three different but conceptually equivalent settings emerged that allow to study the three-body problem in a finite volume: the so-called Relativistic Field Theory (RFT)~\cite{Hansen:2014eka, Hansen:2015zga}, Non-Relativistic Effective Field Theory (NREFT)~\cite{Hammer:2017uqm, Hammer:2017kms} and Finite Volume Unitarity (FVU)~\cite{Mai:2017bge, Mai:2018djl} approaches. Besides this, much work has been done, see, e.g., 
\cite{Meissner:2014dea, Jansen:2015lha, Hansen:2015zta, Hansen:2016fzj, Guo:2016fgl, Konig:2017krd, Briceno:2017tce, Sharpe:2017jej, Guo:2017crd, Guo:2017ism, Meng:2017jgx, Guo:2018ibd, Guo:2018xbv, Klos:2018sen, Briceno:2018mlh, Briceno:2018aml, Doring:2018xxx, Jackura:2019bmu, Mai:2019fba, Blanton:2019igq, Briceno:2019muc, Romero-Lopez:2019qrt, Pang:2019dfe, Guo:2019ogp, Pang:2020pkl, Hansen:2020zhy, Guo:2020spn,Konig:2020lzo,  Blanton:2020gha,Blanton:2020gmf,Brett:2021wyd,Blanton:2021mih,Kreuzer:2010ti, Kreuzer:2009jp, Kreuzer:2008bi, Kreuzer:2012sr,Briceno:2012rv,Blanton:2020jnm}. The finite-volume spectrum has been also
studied in
perturbation theory. In fact, these investigations go back to the 1950's and
have been re-activated recently with the use of the modern technique of the
non-relativistic effective Lagrangians~\cite{Lee:1957zzb, Huang:1957im, Wu:1959zz, Tan:2007bg, Beane:2007qr, Detmold:2008gh, Beane:2020ycc, Pang:2019dfe, Hansen:2016fzj, Romero-Lopez:2020rdq,Muller:2020vtt}. Furthermore,
in quite a few recent papers, the theoretical approaches mentioned above 
have been successfully used to analyze data from lattice calculations~\cite{Beane:2007es, Detmold:2008fn, Detmold:2008yn, Blanton:2019vdk, Horz:2019rrn, Culver:2019vvu, Fischer:2020jzp, Hansen:2020otl, Alexandru:2020xqf, Romero-Lopez:2018rcb, Romero-Lopez:2020rdq,Blanton:2021llb,Brett:2021wyd}. Last but not least,
a three-body analog
of the Lellouch-L\"uscher formula for the finite-volume matrix elements~\cite{Lellouch:2000pv} has been recently derived in two different settings~\cite{Muller:2020wjo,Hansen:2021ofl}. These developments are extensively covered in
the latest reviews on the subject, to which the reader is referred for further details~\cite{Hansen:2019nir,Mai:2021lwb}.

In this paper, we put the issue of the relativistic invariance of the
quantization condition under a detailed scrutiny. The reason for this
is obvious. Typical momenta of light particles (most
notably, pions),
which are studied on the lattice, are not small as compared to their
masses. Albeit the four- and more particle channels might be closed,
or contribute very little, a purely kinematic effect of the relativistic
invariant treatment could be still sizable, especially, if data from the moving frames are considered. For this reason, providing a
manifestly relativistic invariant framework for three particles is extremely
important\footnote{This statement, obviously, refers to the three-particle system in the infinite volume. In a finite volume, the relativistic invariance is anyway broken by the presence of a box.}.

On the other hand, in the derivation of the quantization condition in a field theory one faces a dilemma (this problem concerns all formulations, albeit it is treated differently in different formulations). The amplitudes that enter the quantization condition should be on mass shell -- otherwise, these will not be observables and there will be little use of such a quantization condition. Hence,
the quantization condition is inherently three-dimensional (i.e., it involves
integrations/sums over three-momenta, with the fourth component fixed on mass shell),
and further effort is needed to rewrite it in the manifestly invariant
form.

It is natural to ask the question why the manifest invariance of the setting
is important.
The (three-dimensional) Faddeev equation, which is obeyed by the infinite-volume
amplitude, contains what can be termed as a short-range three-body force
(this quantity enters the finite-volume quantization condition as well,
and its name is different in different approaches).
In principle, choosing this three-body force properly, it should be
always possible to achieve the invariance of the amplitude
(because the true amplitude is invariant and obeys the same equation).
However, implementing this program in practice represents
a very difficult task. Namely, finding an explicit parameterization
of the three-body force that renders the solution
of the Faddeev equations invariant most probably will prove impossible.
Furthermore, making the tree-level
kernel of the Faddeev equation relativistic invariant
order by order in the effective field theory expansion will not suffice
-- without further ado -- to ensure the invariance of the amplitude at the same order,
because the cutoff regularization, which is used in Faddeev equation, breaks
counting rules.
All this results in a very cumbersome and obscure treatment of the problem
that one should better avoid.
On the contrary, in case of a manifestly invariant formulation, the three-body force can be readily parameterized in terms of Lorentz-invariant structures only, see, e.g., a nice discussion in Ref.~\cite{Blanton:2019igq}. The couplings appearing in front of these structures are mutually independent
and the expansion of the short-range part can be organized in accordance with
the well-defined counting rules.
Hence, the advantages of having a manifestly invariant formulation are
evident.

As mentioned above, additional effort is needed to rewrite the three-dimensional
Faddeev equations (infinite volume) and the quantization condition (finite volume) in the manifestly invariant form. As we shall see later,
the problem arises because the three-particle propagator, which originally appears in these equations, is non-invariant.
As a cure to the problem, within the
RFT approach,
it was proposed to modify the three-body propagator, bringing it
to a manifestly invariant form
(the pertinent formulae are given, e.g., in Ref.~\cite{Blanton:2019vdk},
see also Ref.~\cite{Mai:2017vot})
\footnote{Note that the same technique could be used, without any modification, in the NREFT approach as well.}.
We shall briefly consider this prescription below, in Sect.~\ref{sec:3}.
It can be however shown that the
modified propagator breaks unitarity at low energies
(in the infinite volume) and leads to the spurious energy levels below
three-particle threshold (in a finite volume), if the cutoff on the spectator
momentum exceeds some critical value of order of the particle mass itself.
As a result, if one uses the modified propagator, one cannot choose an arbitrarily high cutoff. This is a limitation of the RFT method.

The aim of the present paper is to close the above gap.
Our method is based on the ``covariant'' version of the
NREFT, considered in
Refs.~\cite{Colangelo:2006va,Gasser:2011ju,Bissegger:2008ff,Bissegger:2007yq,Gullstrom:2008sy}. We modify that framework,
choosing the quantization axis along arbitrary timelike unit vector
$v^\mu$ and demonstrate an explicit relativistic invariance of the obtained Faddeev equations with respect to the Lorentz boosts (the original framework corresponded to the choice $v^\mu=v_0^\mu=(1,{\bf 0})$).
The explicit relativistic invariance of the framework emerges if
the vector $v^\mu$ is fixed in terms of the initial and final momenta
in the three-particle system (an obvious choice is to take $v^\mu$
proportional to the total four-momentum).
It is further shown that
there is no restriction on the cutoff within our approach, no breaking of unitarity and no spurious poles for high values of the cutoff.

Last but not least,
we carry out a full group-theoretical analysis of the quantization condition both in the center-of-mass (CM) frame and in moving frames. Namely, the quantization condition is diagonalized into the various
irreducible representations (irreps) of the pertinent point groups. The theoretical constructions are verified numerically, solving the quantization condition for a toy model.

To simplify the argument as much as possible, we consider the case of three identical bosons with a mass $m$ and assume that all Green functions with odd number of external legs identically vanish. These simplifications are of purely technical nature and can be straightforwardly relaxed. The layout of the paper is the following. In Sect.~\ref{sec:2} we consider the two-body sector of the theory and the introduction of auxiliary dimer fields.
The three-body sector is considered in Sect.~\ref{sec:3}, where it is shown that the obtained Faddeev equation for the particle-dimer scattering is explicitly relativistic invariant. In this section, we also
derive the relativistic invariant quantization condition
and carry out a partial diagonalization of this condition into different irreps. In Sect.~\ref{sec:numerics} we investigate the synthetic
three-particle spectrum, obtained in a toy model with the use of the novel
quantization condition. Sect.~\ref{sec:concl} contains our conclusions.

\section{Two-body sector}
\label{sec:2}

\subsection{Threshold expansion}

The non-relativistic approach treats time and space directions differently
that leads to an inherent non-covariance. A trick which allows
one to rewrite all expressions in an explicitly covariant manner is to
introduce an arbitrary unit timelike vector $v^\mu$ and to consider
the time evolution along the axis defined by this vector. The choice
of the ``rest frame'' $v^\mu=v_0^\mu=(1,{\bf 0})$ corresponds to the ``standard'' NREFT.
According to the Lorentz invariance, all choices of $v^\mu$ are
physically equivalent and describe the time evolution as seen by
different moving observers. Note also that a similar trick (albeit in a slightly different physical context) is also used
in the 
Heavy Quark Effective Theory and the Heavy Baryon Chiral Perturbation
Theory.

Of course, indroducing the vector $v^\mu$ alone does not solve the
problem of the non-covariance -- it just allows to recast it fancier.
The presence of an {\em external} vector $v^\mu$ signals
non-covariance. The situation however changes, if it is possible to
express $v^\mu$ through the momenta that characterize a given process.
Then, if the latter are boosted, $v^\mu$ is boosted as well, rendering the amplitudes explicitly Lorentz-covariant. This provides exactly the
solution we are looking for. In the discussion below, we keep $v^\mu$
arbitrary in the beginning, and fix it in terms of the physical momenta
at a later stage.

We start with the construction of the non-relativistic Lagrangians.
In the ``rest frame,'' these are written down, e.g., in
Refs.~\cite{Colangelo:2006va,Gasser:2011ju} on the basis of the
following considerations:
\begin{itemize}
\item
  The nonrelativistic theories do not include the creation and annihilation
  of particles and antiparticles explicitly (the latter can be barred from
  the theory altogether, if one considers the processes with only
  particles in the initial/final states). The effects of creation and
  annihilation are not neglected but consistently included in the
  effective couplings. Hence, the non-relativistic Lagrangian
  is linear in the time derivative, and the propagators feature
  only particle or only antiparticle pole.
\item
  No approximation is made in the energy of the
  free particle $w({\bf k})=\sqrt{m^2+{\bf k}^2}$.
  This ensures that the low-energy singularities of the Feynman diagrams are located exactly at the right place at all orders of the non-relativistic
  expansion.
\item
  The normalization of the non-relativistic field is chosen so that the normalization of the one-particle states in the relativistic and non-relativistic theories is the same.
\end{itemize}
  Below, the Lagrangians derived in Refs.~\cite{Colangelo:2006va,Gasser:2011ju} are rewritten in an arbitrary frame defined by the vector $v^\mu$. The kinetic part then takes the form
\eq
{\cal L}_{\sf kin}=\phi^\dagger 2w_v(i(v\partial)-w_v)\phi\, .
\en
Here, $\phi(x)$ is the non-relativistic field, describing the particle and
$w_v$ denotes the differential operator
\eq
w_v=\sqrt{m^2+\partial^2-(v\partial)^2}\, .
\en
The free non-relativistic propagator is given by
\eq\label{eq:prop}
i\langle 0|T[\phi(x)\phi^\dagger(y)]|0\rangle&=&\int\frac{d^4k}{(2\pi)^4}\,
e^{-ik(x-y)} D(k)\, ,
\nonumber\\[2mm]
D(k)&=&\frac{1}{2w_v( k)(w_v(k)-vk-i\varepsilon)}\, ,
\en
where $w_v( k)=\sqrt{m^2-k^2+(vk)^2}$. In case of $v^\mu=v_0^\mu$,
the above formulae coincide with the ones from Refs.~\cite{Colangelo:2006va,Gasser:2011ju}.

\begin{figure}[t]
  \begin{center}
    \includegraphics*[width=10.cm]{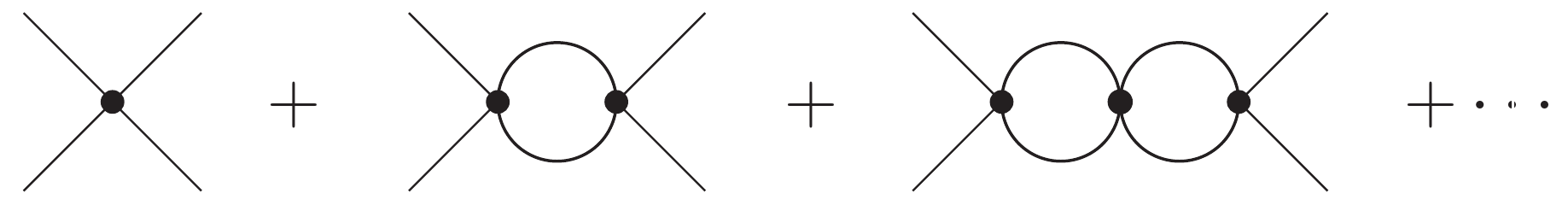}
    \caption{Two-body amplitude in the non-relativistic theory. The filled
    dots denote the full four-particle vertex that can be read off from the interaction Lagrangian. This vertex is a low-energy polynomial.}
    \label{fig:bubbles}
  \end{center}
\end{figure}

Next, let us consider the interactions in the two-particle sector.
The full Lagrangian consists of an infinite tower of terms with
zero, two,\ldots derivatives in the interaction part 
\eq\label{eq:calL}
{\cal L}={\cal L}_{\sf kin}+{\cal L}_0+{\cal L}_2+\cdots\, .
\en\label{eq:calL0}
The lowest-order term is given by
\eq
{\cal L}_0=C_0\phi^\dagger\phi^\dagger\phi\phi\, .
\en
The coupling $C_0$ can be easily related to the two-body S-wave scattering lengths $a_0$ through the matching condition.

As usual, calculating the two-particle scattering amplitude
for the process $p_1+p_2\to q_1+q_2$
with this Lagrangian amounts to summing up all bubble diagrams,
see Fig.~\ref{fig:bubbles}. In the S-wave amplitude, this gives:
\eq
T_0=(4C_0)+(4C_0)^2 \frac{1}{2}\,I+(4C_0)^3\frac{1}{4}\,I^2+\cdots=
\frac{1}{(4C_0)^{-1}-\frac{1}{2}\,I}\, .
\en
Here, $I$ denotes a loop integral
\eq\label{eq:I-infinity}
I=\int\frac{d^Dk}{(2\pi)^Di}\,\frac{1}{2w_v(k)(w_v(k)-vk-i\varepsilon)}\,
\frac{1}{2w_v(P-k)(w_v(P-k)-v(P-k)-i\varepsilon)}\, ,
\en
and $P=p_1+p_2=q_1+q_2$ is the total CM momentum of a particle pair.

Before the evaluation of the above integral the following remarks are in order. First of all, the integrals are ultraviolet-divergent and should be regularized. We use dimensional regularization throughout this paper. This is however not sufficient for ensuring the preservation of counting rules in the loops. To this end, the so-called {\em threshold expansion} (see, e.g.,~\cite{Beneke:1997zp}) should be applied to the loops. One namely first uses the identity
\eq\label{eq:identity}
\frac{1}{2w_v(k)(w_v(k)-vk-i\varepsilon)}
&=&\frac{w_v(k)+vk}{2w_v(k)(m^2-k^2-i\varepsilon)}
\nonumber\\[2mm]
&=&\frac{1}{m^2-k^2-i\varepsilon}
-\frac{1}{2w_v(k)(w_v(k)+vk-i\varepsilon)}\, ,
\en
and a similar identity for the second propagator. Substituting these expression in Eq.~(\ref{eq:I-infinity}), one gets four terms. Furthermore, the threshold expansion is applied in the vicinity of the {\em particle} poles,
$vk=w_v(k)$ and $v(P-k)=w_v(P-k)$, respectively. Moreover,
it is assumed
that the ``three-momenta'' with respect to the quantization axis
$v^\mu$, defined as $k_\perp^\mu=k^\mu-v^\mu vk$
 and $(P-k)_\perp^\mu=(P-k)^\mu-v^\mu v(P-k)$
 are small as compared to the particle mass $m$.\footnote{
   In the standard formulation of the threshold expansion (in the ``rest frame''),
   it is assumed that the
   components of the three-momentum are small,  ${\bf k}^2\ll m^2$.
   More precisely, one introduces a generic small parameter $\epsilon$ and counts
   ${\bf k}=O(\epsilon)$, $k^0=O(1)$.
   Now, assumming (formally) that ${\bf v}=O(\epsilon)$, one immediately sees
   that the components of the vector $k_\perp^\mu=k^\mu-v^\mu vk$ are of order
   $\epsilon$ as well. This counting holds,
   even if $k^\mu$ is an integration momentum. In this case, it is understood merely
   as a prescription that generates threshold expansion in the Feynman integrals.
   It should be further stressed that $\epsilon$ is just a parameter that is used in bookkeeping of various
   contributions. In actual calculations, this parameter may turn out not to be too small.
   A nice example is provided by the three-particle decays of kaons and $\eta$-mesons,
   where the decay products move with the momenta that are not so small as compared to their masses. Despite this fact, the approach works very well~\cite{Colangelo:2006va,Gasser:2011ju}. Note also that the results of the present paper (a derivation of the relativistically invariant
   quantization condition, see below) are exact (to all orders in $\epsilon$) and do not use a
   particular numerical value of $\epsilon$. They simply rely on the fact that one can expand
 the integrand in powers of $\epsilon$, carry out the integration in dimensional regularization and resum the final result again.}
 This means that the second term in the above expression can be expanded as
\eq\label{eq:expanded}
-\frac{1}{2w_v(k)(w_v(k)+vk-i\varepsilon)}
&=&-\frac{1}{4w_v^2(k)}-\frac{w_v(k)-vk}{8w_v^3(k)}+\cdots
\nonumber\\[2mm]
&=&-\frac{1}{4m^2}+\frac{k_\perp^2}{4m^4}+\cdots
-\frac{m-vk}{8m^3}+\cdots\, ,
\en
where the relation $w_v^2(k)=m^2-k_\perp^2$ has been used.
A similar expansion can be written down for the second propagator.
It is now immediately seen that only one term contributes to $I$ after
the threshold expansion since, in the other terms, the integrand becomes
a low-energy polynomial that leads to a vanishing integral in dimensional regularization. Hence, after performing the threshold expansion,
we get:
\eq\label{eq:Is}
I=I(s)=\int\frac{d^Dk}{(2\pi)^Di}\,\frac{1}{(m^2-k^2)(m^2-(P-k)^2)}
=\mbox{const}+\frac{\sigma}{16\pi^2}\,\ln\left(\frac{\sigma-1}{\sigma+1}\right)\, ,
\en
where
\eq
s=P^2\, ,\quad\quad\sigma=\left(1-\frac{4m^2}{s+i\varepsilon}\right)^{1/2}\, .
\en
The renormalization prescription is chosen so that $I(s)$
vanishes at the two-particle threshold $s=4m^2$.

The expression of the loop function, given in Eq.~(\ref{eq:Is}), is explicitly
Lorentz-invariant (depends on the variable $s$ only). It also differs from the expression used in Refs.~\cite{Colangelo:2006va,Gasser:2011ju,Bissegger:2008ff,Bissegger:2007yq,Gullstrom:2008sy}. Namely, the imaginary parts of these two expressions coincide above elastic threshold that ensures two-body unitarity. Moreover, their difference is a low-energy polynomial with real coefficients and, hence, the choice of the loop function
in a form given by Eq.~(\ref{eq:Is}) is as legitimate as the choice made earlier in Refs.~\cite{Colangelo:2006va,Gasser:2011ju,Bissegger:2008ff,Bissegger:2007yq,Gullstrom:2008sy} -- these two correspond to a different renormalization
prescription in the effective theory. Below, we shall stick to the definition
given in Eq.~(\ref{eq:Is}). It has the advantage that the loop function is real and non-singular below threshold, whereas the original definition
leads to a spurious singularity at $s=0$ and to an imaginary part below
this value (we remind the reader that the point $s=0$ lays already outside the region of the applicability of the NREFT, so the question about the
consistency of the approach does not arise here).

Note also that the original derivation given in the above papers was much shorter -- there, one first integrated over the variable $k^0$ and then manipulated the integrand, depending on the three-momenta only. In case of arbitrary $v^\mu$, the dependence of the integrand on $k^0$ is more complicated. In principle, in the infinite volume, one could first perform a Lorentz boost that brings the vector $v^\mu$ to $v_0^\mu$ and then repeat the  steps outlined in these papers. The result will of course be the same. We however stick to this derivation that can be applied in a finite volume without much ado.

In the following, it will be useful to rewrite the loop function as
\eq\label{eq:Js}
I(s)=J(s)+\frac{i\sigma}{16\pi}\, .
\en
Here, as mentioned before, the function $J(s)$ is a low-energy polynomial
with real coefficients.

\subsection{Terms with higher derivatives}

The terms with higher derivatives, present in the Lagrangian, are of two
types. The terms of the first type correspond to the effective-range expansion in a given partial wave (S-wave, in our case), and the terms of a second type describe higher partial waves.

Let us start with the former. The Lagrangian
\eq\label{eq:calL2}
{\cal L}_2=C_2\biggl\{
\bigl((w_\mu\phi)^\dagger (w^\mu\phi)^\dagger\phi\phi-m^2\phi^\dagger\phi^\dagger\phi\phi\bigr)+\mbox{h.c.}\biggr\}
\en
encodes the term related to the $S$-wave effective range $r_0$. Here,
\eq
w^\mu=v^\mu w_v+i\partial_\perp^\mu\, ,\quad\quad
\partial_\perp^\mu=\partial^\mu-v^\mu v\partial\, .
\en
Furthermore, the tree amplitude in a theory consists of the contributions from ${\cal L}_0\,,\,{\cal L}_2\, ,\ldots$, see Eqs.~(\ref{eq:calL}), (\ref{eq:calL0}) and (\ref{eq:calL2}):
\eq
T_{\sf tree}=T^{(0)}_{\sf tree}+T^{(2)}_{\sf tree}+\cdots\, .
\en
As we already know, at lowest order,
\eq
T^{(0)}_{\sf tree}=4C_0\, .
\en
Using now Eq.~(\ref{eq:calL2}), it is straightforward to derive that, on mass shell,
\eq
T^{(2)}_{\sf tree}=4C_2\bigl(s-4m^2\bigr)\, ,
\en
where $s=(\tilde p_1+\tilde p_2)^2=(\tilde q_1+\tilde q_2)^2$, and
$\tilde p_i^\mu=v^\mu w_v(p_i)+p_{i\perp}^\mu$ (similarly for
$\tilde q_i^\mu$ and any other vector).
On the mass shell, where
$w_v(p_i)=vp_i$, $w_v(q_i)=vq_i$
and, consequently,
$\tilde p_i=p_i$, $\tilde q_i=q_i$,
it also follows  that $s=(p_1+p_2)^2=(q_1+q_2)^2$.

It is now crystal clear, how things proceed at higher orders. The
tree-level amplitude in the S-wave represents a Taylor series in $s-4m^2$:
\eq\label{eq:higherorders}
T^{\sf S-wave}_{\sf tree}=4C_0+4C_2(s-4m^2)+4C_4(s-4m^2)^2+\cdots\, ,
\en
All this is fine at tree level, on the mass shell. In the bubble sum, however,
the intermediate particles are off the mass shell. Consider, for example the process $p_1+p_2\to k_1+k_2$, where $p_i^2=m^2$ and $k_i^2\neq m^2$. Then, $s_p=(\tilde p_1+\tilde p_2)^2$ is not equal to
$s_k=(\tilde k_1+\tilde k_2)^2$, even if the relation $p_1+p_2=k_1+k_2$ always holds. Furthermore, the difference between $s_p$ and
$s_k$ is proportional to $w_v(p_1)+w_v(p_2)-w_v(k_1)-w_v(k_2)$. Carrying
out now the contour integration, one can straightforwardly ensure that such a term
in the numerator cancels exactly with the denominator. One is left with a low-energy polynomial, and the integral over this polynomial vanishes in
dimensional regularization. Therefore, replacing $s_k$ by $s_p=s$
everywhere in the numerator is justified. One may finally conclude that
one could consistently pull out the numerator from the integral and evaluate it on shell. This results in the following S-wave amplitude:
\eq\label{eq:T0s}
T_0(s)=\frac{1}{(T^{\sf S-wave}_{\sf tree})^{-1}-\frac{1}{2}\,I(s)}.
\en
Just above threshold, $s>4m^2$, one may rewrite this expression as
\eq
T_0(s)=\frac{16\pi\sqrt{s}}{16\pi\sqrt{s}((T^{\sf S-wave}_{\sf tree})^{-1}-\frac{1}{2}\,J(s))-ip(s)}\, ,\quad\quad p(s)=\sqrt{\frac{s}{4}-m^2}\, .
\en
Thus, just above threshold, the tree amplitude can be related to the
S-wave scattering phase shift
\eq\label{eq:pcotd}
16\pi\sqrt{s}((T^{\sf S-wave}_{\sf tree})^{-1}-\frac{1}{2}\,J(s))=p(s)\cot\delta_0(s)\, .
\en
Expanding both sides in powers of $(s-4m^2)$, one may carry out
the matching of the constants $C_0,C_2,\ldots$ and the effective-range
parameters in the S-wave $a_0,r_0,\ldots$. The lowest-order relation $C_0=-8\pi ma_0$ is the
same as  in Refs.~\cite{Colangelo:2006va,Gasser:2011ju} -- the modifications emerge, starting from the second order only. 

Considering higher partial waves is a bit more subtle because the pertinent amplitudes depend on the directions of momenta as well. In the simple model considered, there is no P-wave. The lowest-order
contribution of the D-wave is
captured by the Lagrangian
\eq
{\cal L}_4&=&{\cal L}_4^{\sf S}+{\cal L}_4^{\sf D}\, ,
\nonumber\\[2mm]
{\cal L}_4^{\sf S}&=&4C_4\biggl(\bigl((w_\mu\phi)^\dagger (w^\mu\phi)^\dagger-m^2\phi^\dagger\phi^\dagger\bigr)\bigl((w_\nu\phi) (w^\nu\phi)-m^2\phi\phi\bigr)\biggr)\, ,
\nonumber\\[2mm]
{\cal L}_4^{\sf D}&=&\frac{5}{2}\,D_4
\biggl(3(w_\mu\phi)^\dagger (w_\nu\phi)^\dagger(w^\mu\phi)(w^\nu\phi)-(w_\mu\phi)^\dagger (w^\mu\phi)^\dagger(w_\nu\phi)(w^\nu\phi)
\nonumber\\[2mm]
&-&\frac{m^2}{2}\,\bigl((w_\mu\phi)^\dagger (w^\mu\phi)^\dagger\phi\phi
+\mbox{h.c.}\bigr)-m^4\phi^\dagger\phi^\dagger\phi\phi
\biggr)\, .
\en
The contribution from ${\cal L}_4^{\sf S}$ in the S-wave amplitude is already shown in Eq.~(\ref{eq:higherorders}). The contribution of the second
term contributes to the on-shell tree amplitude in the D-wave:
\eq\label{eq:Dwave}
T^{\sf D-wave}_{\sf tree}&=&\frac{5}{2}\,D_4\biggl(
6\bigl((\tilde p_1\tilde q_1)(\tilde p_2\tilde q_2)+(\tilde p_1\tilde q_2)(\tilde p_2\tilde q_1)\bigr)-4(\tilde p_1\tilde p_2)(\tilde q_1\tilde q_2)
\nonumber\\[2mm]
&-&2m^2\bigl((\tilde p_1\tilde p_2)+(\tilde q_1\tilde q_2)\bigr)-4m^4
\biggr)+\cdots
\nonumber\\[2mm]
&=&4D_4p^4(s)(2\cdot 2+1)P_2(\cos\theta)+\cdots\, ,
\en
where $P_\ell(\cos\theta)$ stand for the Legendre polynomials, and
\eq
\cos\theta=\frac{t-u}{s-4m^2}\, .
\en
Here $s,t,u$ denote usual Mandelstam variables. Note that we have
already used Lorentz invariance -- the above expression does not depend
on the vector $v^\mu$.

Next, let us consider summing up all bubble diagrams in the D-wave amplitude.
The second iteration, for example, can be written as
\eq
\mbox{second iteration}&=&
\int \frac{d^Dk_1}{(2\pi)^Di}\, \frac{d^Dk_2}{(2\pi)^Di}\,
(2\pi)^D\delta^D(p_1+p_2-k_1-k_2)
\nonumber\\[2mm]
&\times&T^{\sf D-wave}_{\sf tree}(\tilde p_1,\tilde p_2;\tilde k_1,\tilde k_2)
D(k_1)D(k_2)T^{\sf D-wave}_{\sf tree}(\tilde k_1,\tilde k_2;\tilde q_1,\tilde q_2)\, .
\en
Here, $D(k)$ denotes the free propagator, see Eq.~(\ref{eq:prop}).

Furthermore, since $p_i,q_i$ are on the mass shell,
$\tilde p_i^\mu=p_i^\mu$ and $\tilde q_i^\mu=q_i^\mu$.
On the contrary, $\tilde k_i^\mu=k_i^\mu+v^\mu(w_v(k_i)-vk_i)\neq k_i^\mu$.
The additional term cancels with the denominator in $D(k_i)$, leaving
us, after performing the contour integral,
with an integral over the low-energy polynomial that vanishes in the dimensional regularization. Hence, a replacement $\tilde k_i^\mu\to k_i^\mu$
in the numerator is justified. Next, using Eq.~(\ref{eq:identity}), we may rewrite the above equation as:
\eq
\mbox{second iteration}&=&
\int \frac{d^Dk_1}{(2\pi)^Di}\, \frac{d^Dk_2}{(2\pi)^Di}\,
(2\pi)^D\delta^D(p_1+p_2-k_1-k_2)
\nonumber\\[2mm]
&\times&\frac{T^{\sf D-wave}_{\sf tree}(p_1,p_2;k_1,k_2)T^{\sf D-wave}_{\sf tree}(k_1,k_2;q_1,q_2)}
{(m^2-k_1^2-i\varepsilon)(m^2-k_2^2-i\varepsilon)}\, .
\en
It is seen that this integral is written down in a completely
Lorentz-invariant form. In order to evaluate it, we perform the boost
to the center-of-mass frame of two particles.
In this system, the angular integral can be readily done, yielding the Legendre 
polynomial.
Pulling again the numerator out from the integral, we finally get:
\eq
\mbox{second iteration}=(4D_4p^4(s))^2\frac{1}{2}\,I(s)
(2\cdot 2+1)P_2(\cos\theta)\, .
\en
Now, it is easy to write down the result for the D-wave amplitude, summing up the bubbles at all orders:
\eq
T_2(s)=\frac{p^4(s)}{(4D_4)^{-1}-\frac{1}{2}\,p^4(s)I(s)}\, .
\en
Furthermore, it is already clear that, if higher-order terms are taken into account,
the D-wave amplitude takes the form
\eq
T_2(s)=\frac{p^4(s)}{(T^{\sf D-wave}_{\sf tree})^{-1}-\frac{1}{2}\,p^4(s)I(s)}\, ,
\quad\quad
T^{\sf D-wave}_{\sf tree}=4D_4+4D_6(s-4m^2)+\cdots\, .
\en
The matching condition in the D-wave is given by:
\eq
16\pi\sqrt{s}\bigl((T^{\sf D-wave}_{\sf tree})^{-1}-\frac{1}{2}\,p^4(s)J(s)\bigr)
=p^5(s)\cot\delta_2(s)\, .
\en
We are now in a position to write down the expression of the complete amplitude:
\eq\label{eq:Tst}
T(s,t)=\sum_\ell(2\ell+1) P_\ell(\cos\theta)T_\ell(s)\, ,\quad\quad
T_\ell(s)=\frac{p^{2\ell}(s)}{(T^\ell_{\sf tree})^{-1}-\frac{1}{2}\,p^{2\ell}(s)I(s)}\, ,
\en
where $T^\ell_{\sf tree}$ represents a low-energy polynomial in the variable $(s-4m^2)$. The matching condition
\eq
16\pi\sqrt{s}\bigl((T^\ell_{\sf tree})^{-1}-\frac{1}{2}\,p^{2\ell}(s)J(s)\bigr)
=p^{2\ell+1}(s)\cot\delta_\ell(s)
\en
allows one to perform the matching of the couplings in the low-energy effective Lagrangian to the parameters of the effective-range expansion in all partial waves.

In conclusion, we would like to note that, albeit we have started with an
explicitly non-covariant Lagrangian, the physical amplitudes are relativistic invariant, i.e., do not depend on the vector $v^\mu$. This statement is by no means trivial\footnote{In Refs.~\cite{Colangelo:2006va,Gasser:2011ju}, the invariance was demonstrated for a particular choice $v^\mu=v_0^\mu$.}. The relativistic invariance could be achieved, because
a) the interaction Lagrangian of four pions has a particularly simple form
-- it is a bunch of local vertices, and b) the threshold expansion has been applied in the calculation of Feynman integrals. In the three-particle sector,
neither of these conditions hold. The result depends on $v^\mu$ and the relativistic invariance is achieved, when $v^\mu$ is fixed in terms of the external momenta. Below, in Sect.~\ref{sec:3},
we shall consider this issue in detail. 

\subsection{Introducing dimers}

As in Refs.~\cite{Hammer:2017uqm, Hammer:2017kms}, we shall
introduce dimer fields in the Lagrangian in order to trade four-particle
interactions in favor of particle-dimer vertices. This will lead to a
significant simplification in the description of  the three-particle systems,
since the bookkeeping of different diagrams is made much easier in the
particle-dimer picture. Note also that, according to our philosophy,
introducing a dimer does not necessarily mean that a physical dimer
(two-particle bound state) should exist, albeit this may still be the case.
Thus, the particle-dimer formalism is not an {\em approximation}
-- rather, it is a {\em different choice of variables} in the path integral,
equivalent to the original formulation. Note also that, instead of a single
dimer field, we in fact have
to introduce an infinite bunch of dimer fields with different spin, corresponding to different angular momenta $\ell$ in the two-particle system.

Let us again start with the S-wave, and consider the Lagrangian
\eq\label{eq:Ls}
{\cal L}_{\sf S}&=&\phi^\dagger 2w_v(i(v\partial)-w_v)\phi
+\sigma T^\dagger T
\nonumber\\[2mm]
&+&\biggl(\frac{1}{2}\,T^\dagger\bigl(f_0\phi\phi+
f_2\bigl((w_\mu\phi)(w^\mu\phi)-m^2\phi\phi\bigr)+\cdots\bigr)
+\mbox{h.c.}\biggr)+\cdots\, .
\en
Here, $T$ denotes a (scalar) dimer field which does not possess a kinetic
term, and $\sigma=\pm 1$, depending on the sign of the coupling
$C_0$. It is easily seen that, integrating out the dimer field in the path integral, we arrive at the four-particle local coupling one has started with. It is then a simple algebraic exercise to express the new couplings $f_0,f_2,\ldots$ through the $C_0,C_2,\ldots$ and $\sigma$.

The inclusion of the dimers with higher spins proceeds similarly --
one has to merely reformulate the construction of Ref.~\cite{Hammer:2017kms} in the present relativistic setting. To this end, we introduce the
tensor dimer fields $T_{\mu_1,\cdots,\mu_\ell}$, corresponding to the angular
momentum $\ell$. These fields are symmetric under the permutation
of each pair of indices, traceless in each pair of indices
and obey the constraints\footnote{It should be noted out that
  $T_{\mu_1,\cdots,\mu_\ell}$ does not correspond do the standard
  definition of a massive tensor field. For example, a massive vector
  field obeys a constraint $\partial^\mu T_\mu=0$ instead of $v^\mu T_\mu=0$. However, {\em on the mass shell,} these two definitions are related by the
  Lorentz boost that makes the four-momentum of the dimer parallel
to $v^\mu$. }:
\eq\label{eq:vT0}
v^{\mu_i}T_{\mu_1,\cdots,\mu_\ell}=0\, ,\quad\quad i=1,\cdots,\ell\, .
\en
These constraints leave the correct number of independent degrees 
of freedom, equal to $2\ell+1$..
The Lagrangian in the two-particle sector
can be written as:
\eq
{\cal L}=\phi^\dagger 2w_v(i(v\partial)-w_v)\phi
+\sum_{\ell=0}^\infty
\sigma_\ell T^\dagger_{\mu_1,\cdots,\mu_\ell} T^{\mu_1,\cdots,\mu_\ell}
+\frac{1}{2}\,\sum_{\ell=0}^\infty\bigl(T^\dagger_{\mu_1,\cdots,\mu_\ell}O^{\mu_1,\cdots,\mu_\ell}+\mbox{h.c.}\bigr)\, ,\quad\quad
\en
where $\sigma_\ell=\pm 1$ and $O^{\mu_1,\cdots,\mu_\ell}$ are the relativistic two-particle operators, corresponding to the orbital momentum $\ell$.
These operators can be easily constructed, based on the explicit expression of the spherical functions. For example, the lowest-order operator in the D-wave is given by:
\eq
O^{\mu\nu}&=&g_0\biggl(\frac{3}{2}\,(\phi({\bar w}_\perp^\mu {\bar w}_\perp^\nu\phi)
-({\bar w}_\perp^\mu\phi)({\bar w}_\perp^\nu\phi))
\nonumber\\[2mm]
&-&\frac{1}{2}\,(g^{\mu\nu}-v^\mu v^\nu)
(\phi( {\bar w}_\perp^\lambda \bar w_{\perp\lambda}\phi)
-({\bar w}_\perp^\lambda\phi)(\bar w_{\perp\lambda}\phi))\biggr)\, ,
\en
where ${\bar w}_\perp^\mu=\bar w^\mu-v^\mu(v\bar w)$ and $\bar w^\mu$
denotes the operator $w^\mu$ which is boosted in the CM system of two
particles {\em with respect to the vector $v^\mu$.}
Under this, we mean that the boosted total momentum of two particles
on mass shell is parallel to the vector $v^\mu$. Needless to say that, in
a particular case $v^\mu=v_0^\mu$, we get the usual definition of the
two-particle CM frame.

The transformation of $w^\mu$ to $\bar w^\mu$ is given through the matrix
\eq
\bar w^\mu=\Lambda^\mu_\nu w^\nu\, .
\en
It is easier to work in the momentum space. Let $\tilde p_{1,2}$ be the on-mass shell momenta of individual particles. Then, $P=\tilde p_1+\tilde p_2$ is the total on-mass shell momentum of the pair. The boost makes
the vector $P^\mu$ parallel to $v^\mu$, that is\footnote{It is important to mention here that one can {\em always} find such a boost, because both particles are on mass shell, i.e., $P^2\geq 4m^2$.
  This is different, e.g., in the RFT formalism~\cite{Hansen:2014eka, Hansen:2015zga}, where the square of the total momentum can have any sign. However, as mentioned in Refs.~\cite{Wu:2015evh,Li:2021mob,Wu}, there exists an ambiguity in the definition of Lorentz-transformed quantities for the off-shell amplitudes, and the possibility that was described above represents one of the options. In the context of the RFT formalism, this option was explored in detail in Ref.~\cite{Blanton:2020gha}.}
\eq\label{eq:parallel}
\bar P^\mu=\Lambda^\mu_\nu P^\nu=\sqrt{P^2}v^\mu\, ,\quad\quad
{\Lambda^{-1}}^\mu_\nu v^\nu=\frac{1}{\sqrt{P^2}}\,P^\mu\, .
  \en
  This leads to
  \eq
  {\bf v}^2{\bf \bar P}={\bf v}({\bf v}{\bf \bar P})\, ,\quad\quad
  |{\bf v}|\bar P^0=v^0|{\bf\bar P}|\, .
  \en
  The above identities suffice to express the matrix elements of $\Lambda^\mu_\nu$ in terms of the vectors $v^\mu,P^\mu$. Substituting back into the Lagrangian, one should replace the components of $P^\mu$
  by the operators $w^\mu$, acting on the $\phi$ fields.
  The resulting explicit expression is rather voluminous and non-local. It is always implicitly assumed that, in actual calculations, the pertinent expressions are expanded in the inverse powers of the mass $m$, the result is integrated in dimensional regularization and summed up back to all orders.
  Also, we do not display here the explicit expression of the matrix $\Lambda^\mu_\nu$, because it will never be needed.

  In the momentum  space, the lowest-order D-wave
  two-particle-dimer vertex  is given by
\eq
\Gamma^{\mu\nu}(p)=-4g_0\biggl(\frac{3}{2}\,\bar p_\perp^\mu\bar p_\perp^\nu-\frac{1}{2}\,(g^{\mu\nu}-v^\mu v^\nu)(\bar p_\perp)^2\biggr)\, ,
\en
where $\bar p^\mu=\frac{1}{2}\,(\bar p_1^\mu-\bar p_2^\mu)$. Now, integrating out the dimer field $T^{\mu\nu}$, we arrive at
\eq
\Gamma^{\mu\nu}(p)\Gamma_{\mu\nu}(q)
=16g_0^2\biggl(\frac{9}{4}\,(\bar p_\perp\bar q_\perp)^2
-\frac{3}{4}\,(\bar p_\perp)^2(\bar q_\perp)^2\biggr)\, .
\en
Furthermore,
\eq
\bar p_\perp^\mu=\bar p^\mu-v^\mu (v\bar p)
=\bar p^\mu-v^\mu (\Lambda^{-1}v)_\nu p^\nu
=\bar p^\mu-\frac{1}{2\sqrt{P^2}}\,v^\mu((\tilde p_1+\tilde p_2)(\tilde p_1-\tilde p_2))=\bar p^\mu\, .\quad\quad
\en
Using Lorentz invariance, one can transform back to the laboratory frame:
\eq
\Gamma^{\mu\nu}(p)\Gamma_{\mu\nu}(q)
&=&24g_0^2\biggl(\frac{3}{2}\,(pq)^2-\frac{1}{2}\,p^2q^2\biggr)
=\frac{3}{2}\,
g_0^2\biggl(\frac{3}{2}\,(t-u)^2-\frac{1}{2}\,(s-4m^2)^2\biggr)
\nonumber\\[2mm]
&=&24g_0^2p^4(s)P_2(\cos\theta)\, .
\en
This result is similar to Eq.~(\ref{eq:Dwave}) and gives a matching
condition for the variable $g_0$. Inclusions of higher orders in the effective-range expansion, as well as higher partial waves is now straightforward and will not be written down in detail. The only difference to the
``conventional'' case with $v^\mu=v_0^\mu$ is that all momenta are boosted to the CM frame with respect to $v^\mu$, i.e., the total momentum is parallel to $v^\mu$ after the boost. Further, instead of three-momenta
in the boosted frame, the transverse components $p_\perp$ are considered, and the covariant expression $v^\mu v^\nu-g^{\mu\nu}$ replaces the
three-dimensional Kronecker delta in the boosted frame. Last but not least, we wish to reiterate that, unlike the
original formulation of the RFT formalism,  the boost is always well defined in the NREFT framework.
This happens because we work with the on-shell particles. 

Finally note that the two-body on-shell amplitude, given in Eq.~(\ref{eq:Tst}) can be rewritten in the following form
\eq
T(s,t)=4\pi\sum_{\ell m}
\mathscr{Y}_{\ell m}({\bf \tilde{p}})
\frac{1}{(T^\ell_{\sf tree}(s))^{-1}-\frac{1}{2}\,p^{2\ell}(s)I(s)}\,
\mathscr{Y}^*_{\ell m}({\bf \tilde{q}})
\, ,
\en
where
\eq
   {\bf \tilde p}={\bf \bar p}-{\bf v}
   \frac{{\bf \bar p}{\bf v}}{{\bf v}^2}+
   {\bf v}\frac{\bar p^0}{{\bf v}^2}\, ,\quad\quad
      {\bf \tilde q}={\bf \bar q}-{\bf v}\frac{{\bf \bar q}{\bf v}}{{\bf v}^2}+
      {\bf v}\frac{\bar q^0}{{\bf v}^2}\, ,\quad\quad
 {\bf \tilde p}    {\bf \tilde q}=-\bar p_\mu\bar q^\mu\, ,    
      \en
      and
      \eq
      \mathscr{Y}_{\ell m}({\bf k})=|{\bf k}|^\ell Y_{\ell m}(\hat k)\, ,\quad\quad
     \hat k=\frac{\bf k}{|{\bf k}|}\, .
      \en
      Note that, in case of $v^\mu=v_0^\mu$, the above definition of
      the boosted amplitude coincides
      with the boost introduced in~\cite{Wu:2015evh,Li:2021mob,Wu}.
      Within this prescription the boosted three-momenta are always
      well-defined, even for $s<0$.

\section{Three-body sector}
\label{sec:3}

\subsection{Particle-dimer Lagrangian}

The construction of the particle-dimer Lagrangian that describes short-range three-particle
interactions, proceeds analogously to the case of the two-particle Lagrangian, except
three differences. First, a dimer and $\phi$ are not identical particles and hence
all (not only even) partial waves are allowed. Second, in difference with $\phi$,
dimers have spin. And third, in the particle-dimer system one cannot use equations of
motion in order to reduce the number of the independent terms in the Lagrangian. 
The dimers, in general, are unphysical ``particles'' and do not have a fixed mass.

Let us again start with a scalar dimer. The tree-level particle-dimer
scattering amplitude depends on the following kinematic variables:
\eq
s=(p+P)^2=(q+Q)^2\, ,\quad
t=(p-q)^2=(P-Q)^2\, ,\quad
\sigma_p^2=P^2\, ,\quad
\sigma_q^2=Q^2\, ,
\en
where $p,q$ and $P,Q$ are the momenta of incoming/outgoing particles and
incoming/outgoing dimers, respectively. Consistent counting rules can be
imposed, for example, assuming:
\eq
\Delta&=&s-9m^2=O(\epsilon^2)\, ,\quad\quad
t=O(\epsilon^2)\, ,
\nonumber\\[2mm]
\Delta_p&=&\sigma_p^2-4m^2=O(\epsilon^2)\, ,\quad\quad
\Delta_q=\sigma_q^2-4m^2=O(\epsilon^2)\, ,
\en
where $\epsilon$ is a generic small parameter, and all transverse
momenta count as $p_\perp=O(\epsilon)$.

Expanding the tree amplitude in Taylor series, we get:
\eq\label{eq:Td}
T_d^{\sf tree}(s,t,\sigma_p^2,\sigma_q^2)=x_0+x_1(s-9m^2)+x_2t+x_3(\sigma_p^2+\sigma_q^2-8m^2)
+O(\epsilon^4)\, .
\en
Here, we have additionally used the invariance under time reversal that implies the
interchange of the initial and final momenta. In the tree-level amplitude
all coefficients $x_0,x_1,\ldots$ are real due to unitarity.

Furthermore, the couplings $x_0,x_1,\ldots,$ {\em determined from the tree-level matching,} are not all independent. Indeed,
the particle-dimer Lagrangian is used to calculate the three-particle amplitude,
and the matching is performed for the latter. The couplings (or the linear combinations thereof), which do not contribute to the on-shell three-particle amplitude, are redundant and can be dropped.
In order to obtain the three-particle scattering amplitude from
the particle-dimer scattering amplitude, one has to equip the external
dimer legs with two-particle-dimer vertices and sum up over all permutations
in the initial as well as final state. At order $\epsilon^2$, it suffices
to consider the vertex $f=f_0+\frac{1}{2}\,f_2(\sigma^2-4m^2)$, see
Eq.~(\ref{eq:Ls}). Here, $\sigma^2$ stands for the four-momentum square of a dimer. Since any of the initial or final particles can be a spectator,
one has to equip the quantities $\Delta_{p,q}$ and $t$ by
indices $i,j=1,2,3$ that label spectator particles, and sum over these indices.
Thus, one has to define:
\eq
\Delta_p^i=P_i^2-4m^2\, ,\quad\quad
\Delta_q^i=Q_i^2-4m^2\, ,\quad\quad
t^{ij}=(p_i-q_j)^2\, .
\en
These obey the following kinematic identities on mass shell,
see also Ref.~\cite{Blanton:2019igq}:
\eq
\sum_{i=1}^3\Delta_p^i=\sum_{i=1}^3\Delta_q^i=\Delta\, ,\quad\quad
\sum_{j=1}^3t^{ij}=\Delta_p^i-\Delta\, ,\quad\quad
\sum_{i=1}^3t^{ij}=\Delta_q^j-\Delta\, .
\en
The three-particle amplitude is given by
\eq
T_3^{\sf tree}=\sum_{i,j=1}^3f({\sigma_p^i}^2) T_d^{\sf tree}(s,t^{ij},{\sigma_p^i}^2,{\sigma_q^j}^2)
f({\sigma_q^j}^2)+O(\epsilon^4)\, .
\en
Taking into account the above identities, it is straightforward to ensure
that only two independent terms survive in the tree-level three-particle amplitude
at this order:
\eq
T_3^{\sf tree}=z_0+z_1\Delta+O(\epsilon^2)\, .
\en
This agrees with the result of  Ref.~\cite{Blanton:2019igq}.
Moreover, as shown in~\cite{Bedaque:2002yg}, in the particle-dimer
formalism
it is possible to trade the terms of the type $\Delta_p+\Delta_q$
and $\Delta$ for each other\footnote{Ref.~\cite{Bedaque:2002yg} considers the non-relativistic limit and the CM frame only. Hence, strictly speaking,
  this paper discusses the elimination of the next-to-leading contact interaction, proportional to ${\bf p}^2+{\bf q}^2$, in favor of the linear function of the total CM energy $E$.}. Thus our result confirms
the findings of
Ref.~\cite{Bedaque:2002yg} as well. To summarize, only one coupling
out of $x_1,x_2,x_3$ is independent and, without the loss of generality,
one may assume, say, $x_2=x_3=0$ (Note also that
in Refs.~\cite{Hammer:2017uqm,Hammer:2017kms} we have written down an energy-independent next-to-leading order driving term containing ${\bf p}^2+{\bf q}^2$. In the present context, it corresponds to the choice $x_1=x_2=0$ and $x_3\neq 0$.).
Finally, note that a similar analysis can be
carried out at higher orders. We do not consider here this rather
straightforward exercise which, at order $\epsilon^4$, again reproduces
the result of Ref.~\cite{Blanton:2019igq}.

Here one should however note that all the above analysis was limited to the
case when a physical dimer does not exist. In case this is not true, the following line of reasoning can be applied. Let us go back to Eq.~(\ref{eq:Td}). In this case, $\sigma_p^2$ and $\sigma_q^2$ are not independent kinematic variables
anymore, being fixed to the dimer mass squared. On the contrary, the derivative
couplings $x_1,x_2$ can be independently matched to the S-wave effective range
and the P-wave scattering length of the particle-dimer scattering.

Furthermore, note that all discussions up to now
were restricted to the tree
level. Owing to the fact that the use of the cutoff regularization in the
Faddeev equation leads to the breakdown of naive counting rules that
can be rectified only by adjusting the renormalization prescription,
studying the independence of $x_1,x_2,x_3$ in general is a more subtle
issue. In this case, we find it safe to include all couplings -- after all,
using (possibly) an overcomplete set of operators in the Lagrangian
is certainly not a mistake.

A final remark here concerns the situation, where the low-lying three-particle
resonances exist. In this case, the assumption that the short-range part
of the particle-dimer interaction is a low-energy polynomial in $s-9m^2$
might prove to be too restrictive, since the pertinent expansion has
a very small radius of convergence, caused by a nearby resonance. A Laurent
expansion of the short-range interaction, featuring a simple pole
$\sim(s-s_0)^{-1}$ with an unknown parameter $s_0$, describes the system
in a more adequate fashion in this case.

Next, let us briefly dwell on the partial-wave expansion of the particle-dimer
short-range tree amplitude. As seen,
the $O(\epsilon^2)$ amplitude contains only an S-wave contribution. At higher
orders, one can define the scattering angle $\theta$, according to
\eq
t-2m^2=\frac{(s+m^2-\sigma_p^2)(s+m^2-\sigma_q^2)}{4s}
-\frac{\lambda^{1/2}(s,m^2,\sigma_p^2)\lambda^{1/2}(s,m^2,\sigma_q^2)}{4s}\,
\cos\theta\, .
\en
Then, the expansion of the tree particle-dimer
amplitude in the series of Legendre polynomials
can be written down straightforwardly. Note that, at a given order in $\epsilon$, this expansion always contains a finite number of Legendre polynomials.

Having considered the scattering of a particle and a scalar dimer in a
great detail, we now sketch the construction in case of a dimer with arbitrary
integer spin. To this end, it is convenient to use a different basis
for the dimer fields $T_{\mu_1,\cdots,\mu_\ell}$, removing the redundant components.
In order to achieve this, consider first the Lorentz transformation\footnote{Note that this
transformation is different from $\Lambda^\mu_\nu$, considered in the previous section.}
\eq
\underline{\Lambda}^\mu_\nu v^\nu=v_0^\mu\, ,\quad\quad
\underline{T}_{\mu_1,\cdots,\mu_\ell}=\underline{\Lambda}_{\mu_1}^{\nu_1}\cdots\underline{\Lambda}_{\mu_\ell}^{\nu_\ell}T_{\nu_1,\cdots,\nu_\ell}\, .
\en
The transformed field
$\underline{T}$ is zero, if one of the indices $\mu_1,\ldots,\mu_\ell$
is equal to zero, see Eq.~(\ref{eq:vT0}). The space components can be directly related
to the dimer field components $T_{\ell m}$ with $m=-\ell,\ldots, \ell$:
 \eq
 \underline{T}_{\mu_1,\cdots,\mu_\ell}=\sum_{m=-\ell}^\ell c_{\mu_1,\cdots,\mu_\ell}^{\ell m}T_{\ell m}\, .
 \en
 The coefficients $c_{\mu_1,\cdots,\mu_\ell}^{\ell m}$ are pure numbers and can be read off from the explicit expressions of the spherical functions. They are zero, if one of the $\mu_i$ is equal to zero.

 A generic matrix element can be also boosted to the rest frame:
 \eq
 &&\langle p,(P,\mu_1',\cdots,\mu_{\ell'}')|T_d^{\sf tree}|q,(Q,\mu_1,\cdots,\mu_\ell)\rangle=
 (\underline{\Lambda}^{-1})_{\mu_1'}^{\nu_1'}\cdots
 (\underline{\Lambda}^{-1})_{\mu_{\ell'}'}^{\nu_{\ell'}'}
 (\underline{\Lambda}^{-1})_{\mu_1}^{\nu_1}\cdots
 (\underline{\Lambda}^{-1})_{\mu_{\ell}}^{\nu_{\ell}}
\nonumber\\[2mm]
 &&\quad\quad\times\,\sum_{m'=-\ell'}^{\ell'}\sum_{m=-\ell}^{\ell}
 c_{\nu_1',\cdots,\nu_{\ell'}'}^{\ell' m'}c_{\nu_1,\cdots,\nu_\ell}^{\ell m}
 \langle\underline{p},(\ell'm')|T_d^{\sf tree}|\underline{q},(\ell m)\rangle\, ,
 \en
where $\underline{p}^\mu,\underline{q}^\mu$ are the Lorentz-transformed momenta:
\eq
\underline{p}^\mu=\underline{\Lambda}^\mu_\nu p^\nu\, ,\quad\quad
\underline{q}^\mu=\underline{\Lambda}^\mu_\nu q^\nu\, ,
\en
and we anticipated that the total momentum of the system $K^\mu$ is proportional to $v^\mu$, so that the
same Lorentz boost brings the considered matrix element to the
CM frame.

The matrix element in the right-hand side can be expanded in partial waves:
\eq
\langle\underline{p},(\ell'm')|T_d^{\sf tree}|\underline{q},(\ell m)\rangle
=4\pi\sum_{JM}\sum_{L'L}\mathscr{Y}^{L'\ell'}_{JM}({\bf \underline{p}},m')
T_{JL'L}(\Delta,\Delta_p,\Delta_q)
\bigl[\mathscr{Y}^{L\ell}_{JM}({\bf \underline{q}},m)\bigr]^*\, .
\en
Here,
\eq
\mathscr{Y}^{L\ell}_{JM}({\bf \underline{k}},m)
=\langle L(M-m),\ell m|JM\rangle\mathscr{Y}_{L(M-m)}({\bf \underline{k}})
\en
is the spherical function with spin $\ell$, which is given by
an ordinary spherical function multiplied with the pertinent Clebsch-Gordan
coefficient. Note that, for a generic $v^\mu$, the above expansion
has a more complicated form, since an additional boost is needed
to bring the matrix element to the CM frame first\footnote{The pertinent
  boost is given by $U(\tilde \Lambda)|P\ell m\rangle
  =\sum\limits_{m'=-\ell}^\ell D^{(\ell)}_{m'm}(W(\tilde\Lambda,P))|(\tilde\Lambda P)\ell m'\rangle$, where $W(\tilde\Lambda,P)$ denotes the corresponding
  Wigner rotation and $\tilde\Lambda $ is the transformation that brings the particle-dimer system to the rest frame.}. Also,
the above expressions show that in the partial-wave expansion of the
three-particle amplitude one encounters two orbital momenta:
1) the orbital momentum of pairs which in the particle-dimer approach are
represented by the dimer spin $\ell$, and 2) the orbital momentum between
a pair and a spectator, which corresponds to the quantum number $L$. The introduction of dimers allows one
to neatly separate the partial-wave expansion in these two orbital momenta.
The quantity $J$ corresponds to a sum of these orbital momenta and is conserved.

Furthermore, the quantities
$T_{JL'L}(\Delta,\Delta_p,\Delta_q)$  are the low-energy polynomials\footnote{As already mentioned, the low-lying three-body resonances may lead
to the poles in the variable $\Delta$.},
expanded up to a given order in $\epsilon$. 
Like in the case of a scalar dimer, some on-shell constraints
will emerge between various low-energy couplings at a given order in $\epsilon$.
We shall make no attempt here to write down these constraints in a general form.
When needed, this can be most easily done on the case-by-case basis.

A further remark is due at this place, concerning the expression of the
most general Lorentz-invariant short-range amplitude. Namely, in the
construction of the invariant kinematic structures we have never used the vector
$v^\mu$ which should be also included on general grounds.
The excuse is provided by the fact that,
at the end, we shall relate $v^\mu$
to the external momenta (in particular,
we shall take it parallel to the total momentum of the three-particle system).
In this case, all invariants that can be constructed with the use of $v^\mu$ can
be expressed in terms of the already considered ones. Anticipating this fact,
we did not write down such invariants at all.

This concludes the construction of a short-range tree-level
particle-dimer scattering amplitude with initial and final dimers having
any spins $\ell,\ell'$. Construction of such an amplitude is equivalent to the
construction of the particle-dimer Lagrangian. We do not make an attempt to
display such a Lagrangian explicitly, because it is far more convenient
to work directly with the momentum space amplitudes.

\subsection{Faddeev equation for the particle-dimer amplitude}

Now, we are ready to write down the Faddeev equation, describing the particle-dimer scattering, in an explicitly Lorentz-invariant form. In order to avoid cumbersome expressions that will only render the basic idea obscure, we shall first
restrict ourselves to the S-wave interactions in both orbital momenta.
As follows from the discussion above, including higher partial waves merely
amounts to adding indices to some of the quantities in
the expressions. This procedure can be readily carried out.

Let us start from the scalar dimer propagator:
\eq\label{eq:S}
i\langle 0|T[T(x)T^\dagger(y)]|0\rangle=\int\frac{d^4P}{(2\pi)^4}\,e^{-iP(x-y)}S(P^2)\, ,
\en
where
\eq
S(P^2)=\biggl(-\frac{1}{\sigma}\biggr)+\biggl(-\frac{1}{\sigma}\biggr)^2\Sigma(P^2)+\cdots=-\frac{1}{\sigma+\Sigma(P^2)}\, ,
\en
and
\eq
\Sigma(P^2)=\int\frac{d^Dk}{(2\pi)^Di}\,\frac{\frac{1}{2}\,f^2(\bar P^2)}{2w_v(k)(w_v(k)-vk-i\varepsilon)\,2w_v(P-k)(w_v(P-k)-v(P-k)+i\varepsilon)}\, ,
\nonumber\\
\en
with
\eq
\bar P^2=(w_v(k)+w_v(P-k))^2+P_\perp^2\, ,\quad\quad
f(u)=f_0+\frac{1}{2}\,f_2(u-4m^2)+\cdots\, .
\en
Performing the threshold expansion and evaluating the expression in dimensional regularization,
one gets: 
\eq
\Sigma=\frac{1}{2}\,f^2(P^2)I(P^2)\, ,
\en
where $I(P^2)$ is given in Eq.~(\ref{eq:Is}).

\begin{figure}[t]
\begin{center}
\includegraphics*[width=8.cm]{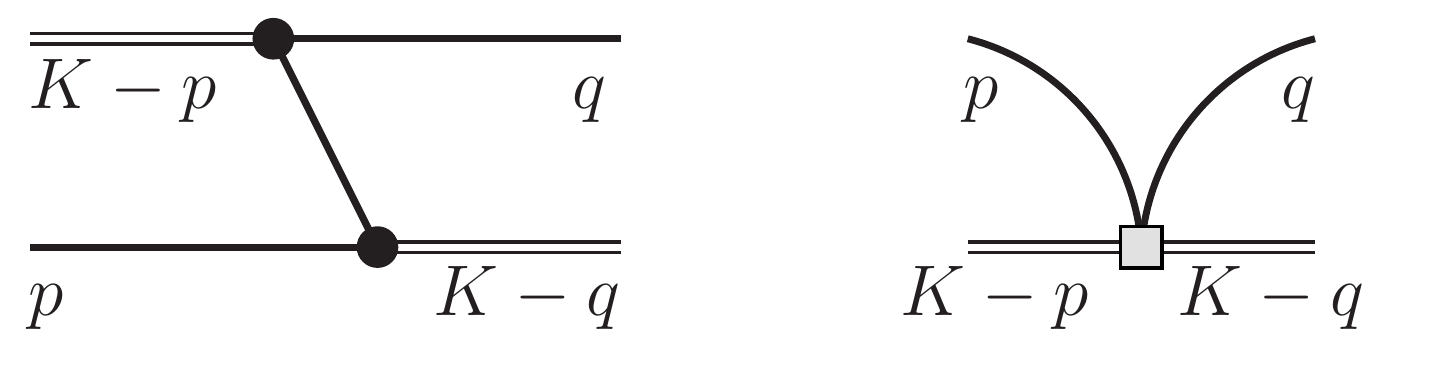}
\caption{The kernel of the Faddeev equation. Double and single lines denote
the dimers and particles, respectively. The dark blob describes a dimer-two-particle vertex, and a shaded box corresponds to the short-range part of the particle-dimer scattering.}
\label{fig:Z}
\end{center}
\end{figure}

Next, let us consider the tree-level particle-dimer scattering amplitude,
which consists of two diagrams shown in Fig.~\ref{fig:Z}. These are diagrams
describing one-particle exchange and the local particle-dimer
interaction. Furthermore, vertices in each diagram consist of an
infinite number of terms, corresponding to the derivative operators in the
Lagrangian. Thus, the tree-level amplitude is given by\footnote{In the
  ``rest system'' $v^\mu=v_0^\mu$, this expression can be obtained
  with the use of the time-ordered perturbation theory. For arbitrary $v^\mu$, one considers instead the evolution in direction of the vector $v^\mu$. The role of the Hamiltonian in this case is played by $\mathbb{H}=v_\mu\mathbb{P}^\mu$, where  $\mathbb{P}^\mu$ denotes the
  operator of the full four-momentum. The four-momentum of a free particle obeys the mass-shell condition $vk=w_v(k)$. It is then clear that, in the frame defined by the vector $v^\mu$, the one-particle exchange diagram takes the form given in Eq.~(\ref{eq:Ttree}). An alternative derivation
  of the same expression starts from the Bethe-Salpeter equation and performs the ``equal-time projection'' of this equation by integrating over
the component of the relative momentum, parallel to the vector $v^\mu$.}
\eq\label{eq:Ttree}
T^{\sf tree}=\frac{f(s_p)f(s_q)}{2w_v(K-p-q)(w_v(p)+w_v(q)+w_v(K-p-q)-vK-i\varepsilon)}+T_d^{\sf tree}\, .
\en
Here, $p,q$ are the four-momenta of the external particles, and $K$ is a total
momentum of a particle-dimer pair. Hence, the four momenta of dimers are
$P=K-p$ and $Q=K-q$. The short-range particle-dimer amplitude is given by
Eq.~(\ref{eq:Td}). Furthermore, the kinematic variables $s_p,s_q$ are given by
\eq
s_p&=&(w_v(p)+w_v(K-p-q))^2+(K-q)_\perp^2\, ,
\nonumber\\[2mm]
s_q&=&(w_v(q)+w_v(K-p-q))^2+(K-p)_\perp^2\, ,
\en
and, for any vector $a^\mu$, we have $a^\mu_\perp=a^\mu-v^\mu(va)$.

Furthermore, let us consider the difference:
\eq
&&s_p-\sigma_q^2=s_p-(K-q)^2=(w_v(p)+w_v(K-p-q))^2-(w_v(p)-v(K-p-q))^2
\nonumber\\[2mm]
&=&(w_v(p)+w_v(K-p-q)+w_v(q)-vK)(w_v(p)+w_v(K-p-q)-w_v(q)+vK)\, .\quad\quad
\en
A similar relation holds for the $s_q-\sigma_p^2=s_q-(K-p)^2$. Taking into account the fact that the function $f(u)$ is a low-energy polynomial in the variable
$u-4m^2$, it is seen that the arguments $s_p,s_q$ in these functions can be
replaced by $\sigma_q^2,\sigma_p^2$. In the difference, the denominator cancels
and hence, it only modifies the regular part $T_d^{\sf tree}$. The fact that the
modified short-range part now depends on the vector $v^\mu$, does not lead
to any problem. One could merely
ignore such $v$-dependent terms since, at the end,
$v^\mu$ will be chosen proportional to $K^\mu$. Thus, one could write
\eq\label{eq:Ttildetree}
\tilde T^{\sf tree}=\frac{f(\sigma_p^2)f(\sigma_q^2)}{2w_v(K-p-q)(w_v(p)+w_v(q)+w_v(K-p-q)-vK-i\varepsilon)}+\tilde T_d^{\sf tree}\, .
\en
Note that the separate terms in Eqs.~(\ref{eq:Ttree}) and (\ref{eq:Ttildetree})
are manifestly invariant, {\em if the vector $v^\mu$
  is also boosted along with all other vectors.} This is different from the standard formulation, where $v^\mu$ is chosen along $v_0^\mu$ and {\em does not transform} under Lorentz transformations.

\begin{figure}[t]
  \begin{center}
    \includegraphics*[width=12.cm]{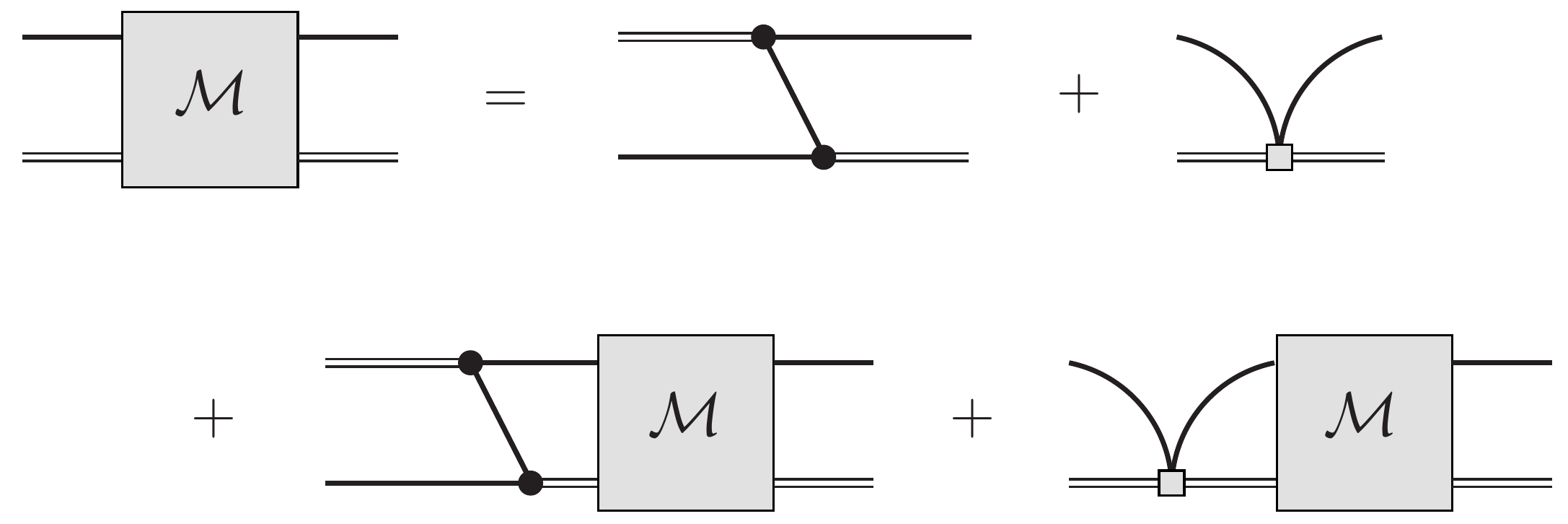}
    \caption{Faddeev equation for the particle-dimer scattering amplitude.}
    \label{fig:Faddeev}
  \end{center}
\end{figure}

Graphically, the Faddeev equation for the particle-dimer scattering amplitude
is depicted in Fig.~\ref{fig:Faddeev}. Denoting this amplitude by $\tilde T$, we have:
\eq\label{eq:tildeT}
\tilde T(p,q)=\tilde T^{\sf tree}(p,q)
+\int^{\Lambda_v}\frac{d^3k_\perp}{(2\pi)^32w_v(k)}\,
\tilde T^{\sf tree}(p,k)S((K-k)^2)\tilde T(k,q)\, ,
\en
where
\eq
\int^{\Lambda_v}\frac{d^3k_\perp}{(2\pi)^32w_v(k)}\,F(k)
=\int\frac{d^4k}{(2\pi)^3}\,\delta(k^2-m^2)\theta(\Lambda^2+k^2-(vk)^2)F(k)\, .
\en
Defining now
\eq
\tilde T(p,q)&=&f(\sigma_p^2){\cal M}(p,q)f(\sigma_q^2)\, ,
\nonumber\\[2mm]
\tilde T^{\sf tree}(p,q)&=&f(\sigma_p^2){\cal Z}(p,q)f(\sigma_q^2)\, ,
\en
we may finally rewrite the Faddeev equation as
\eq
{\cal M}(p,q)={\cal Z}(p,q)
+\int^{\Lambda_v}\frac{d^3k_\perp}{(2\pi)^32w_v(k)}\,
{\cal Z}(p,k)\tau((K-k)^2){\cal M}(k,q)\, .
\en
Here, $\tau(z)$ is the physical two-body scattering matrix
\eq
\tau(z)=f^2(z)S(z)=\frac{1}{-\sigma f^{-2}(z)-\frac{1}{2}\,I(z)}=T_0(z)\, ,
\en
with $T_0(z)$ defined in Eq.~(\ref{eq:T0s}).

\begin{figure}[t]
  \begin{center}
    \includegraphics*[width=12.cm]{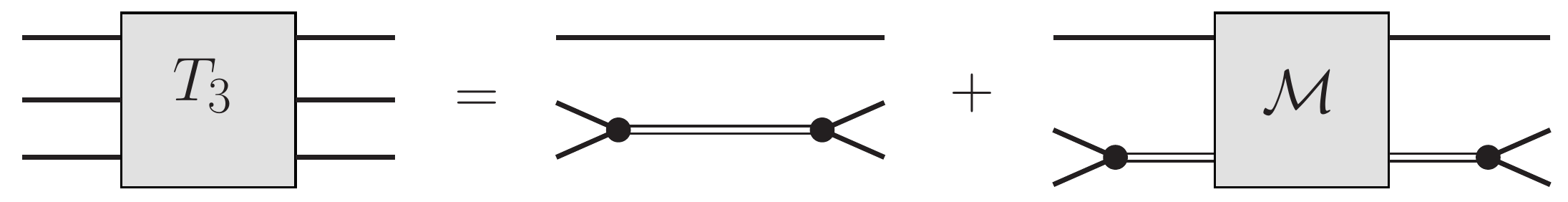}
    \caption{Expressing three-particle amplitude in terms of a particle-dimer amplitude. Summing up over all possible choices of spectator particles is implicit.}
    \label{fig:dimer-3particle}
  \end{center}
\end{figure}

The three-particle amplitude can be expressed through the particle-dimer
amplitude
\eq\label{eq:T3M}
T_3(p_1,p_2,p_3;q_1,q_2,q_3)&=&T_3^{\sf disc}+T_3^{\sf conn}\, ,
\nonumber\\[2mm]
T_3^{\sf disc}&=&\sum_{i,j=1}^3
(2\pi)^3\delta^3(p_{i\perp}-q_{j\perp})2w_v(p_i)
\tau((K-p_i)^2)\, ,
\nonumber\\[2mm]
T_3^{\sf conn}&=&\sum_{i,j=1}^3\tau((K-p_i)^2){\cal M}(p_i,q_j)\tau((K-q_j)^2)\, .
\en
Symbolically, this relation is depicted in Fig.~\ref{fig:dimer-3particle}.

Up to this point, all expressions are manifestly Lorentz invariant, if the
vector $v^\mu$ is also Lorentz-boosted along with other vectors. This means
that, for instance, the particle-dimer amplitude ${\cal M}$, which implicitly depends on
the choice of the quantization axis $v^\mu$, is invariant under arbitrary
Lorentz boosts
\eq
{\cal M}(\Lambda p;\Lambda q;\Lambda v)={\cal M}(p;q;v)\, .
\en
In other words, after fixing $v^\mu$ in terms of the external momenta
$p_i,q_j$ (the most natural choice is, as already mentioned above, to choose
$v^\mu$ along the total four-momentum $K^\mu$ of the three-particle system),
the particle-dimer amplitude becomes manifestly Lorentz-invariant. Thus, the
goal stated in the beginning has been achieved. We would like to stress
here that this happens because the two-particle scattering amplitude after
using the threshold expansion depends only on the pertinent Mandelstam
variable $s$ for a given subsystem and does
not depend on $v^\mu$. If it were not the case, one would be forced to fix
the direction of the quantization axis for each subsystem separately, as
well as for the whole system, and this cannot be done simultaneously.
It is also clear that this approach will face difficulties in the study of the
four-particle system, which contains different
three-particle subsystems.

In conclusion, note that if the dimers with higher spin are taken into account, both the dimer propagator $S$ in Eq.~(\ref{eq:S}) and the tree-level
amplitude $T^{\sf tree}$ become matrices in the space of Lorentz indices, e.g.,
$S\to S_{\mu_1\ldots\mu_{\ell'},\nu_1\ldots\nu_\ell}\doteq S_{n_{\ell'}n_\ell}$ and
$T^{\sf tree}\to T^{\sf tree}_{\mu_1\ldots\mu_{\ell'},\nu_1\ldots\nu_\ell}\doteq T^{\sf tree}_{n_{\ell'}n_\ell}$.
All further steps can be performed in a direct analogy to the case
of the scalar dimer.
Namely, replacing $s_p$ by $\sigma_q^2$ and
$s_q$ by $\sigma_p^2$ is straightforward.
This leads to a system of equations (cf. with Eq.~(\ref{eq:tildeT})):
\eq
\tilde T_{n_{\ell'}n_\ell}(p,q)&=&\tilde T^{\sf tree}_{n_{\ell'}n_\ell}(p,q)
\nonumber\\[2mm]
&+&\sum_{n_{\ell''}n_{\ell'''}}\int^{\Lambda_v}\frac{d^3k_\perp}{(2\pi)^32w_v(k)}\,
\tilde T^{\sf tree}_{n_{\ell'}n_{\ell''}}(p,k)S_{n_{\ell''}n_{\ell'''}}((K-k)^2)\tilde T_{n_{\ell'''}n_\ell}(k,q)\, .\quad\quad
\en
Note that the matrix $S$ is diagonal in $\ell',\ell$ in the infinite, but not
in a finite volume. Furthermore,
\eq
\sum_{n_\ell}(\cdots)=\sum_{\ell=0}^{\ell_{max}}\sum_{\mu_1,\cdots,\mu_\ell}(\cdots)\, .
\en
Next, one may define:
\eq
\tilde T_{n_{\ell'}n_\ell}(p,q)&=&f_{\ell'}(\sigma_p^2){\cal M}_{n_{\ell'}n_\ell}(p,q)
f_\ell(\sigma_q^2)\, ,
\nonumber\\[2mm]
\tilde T^{\sf tree}_{n_{\ell'}n_\ell}(p,q)&=&f_{\ell'}(\sigma_p^2){\cal Z}_{n_{\ell'}n_\ell}(p,q)
f_\ell(\sigma_q^2)\, ,
\nonumber\\[2mm]
\tau_{n_{\ell'}n_\ell}((K-k)^2)&=&f_{\ell'}(\sigma_k^2)S_{n_{\ell'}n_{\ell}}((K-k)^2)f_\ell(\sigma_k^2)\, ,
\nonumber\\[2mm]
\sigma_k^2&=&(K-k)^2-4m^2\, .\quad
\en
In the infinite volume, the matrix $\tau$ is also diagonal and contains
the on-shell two-body partial-wave amplitudes.

The three-body amplitude is given by (cf. with Eq.~(\ref{eq:T3M})
\eq
T_3(p_1,p_2,p_3;q_1,q_2,q_3)&=&T_3^{\sf disc}+T_3^{\sf conn}\, ,
\nonumber\\[2mm]
T_3^{\sf disc}&=&\sum_{i,j=1}^3
(2\pi)^3\delta^3(p_{i\perp}-q_{j\perp})2w_v(p_i)
\nonumber\\[2mm]
&\times&\sum_{n_{\ell'}n_\ell}Y_{n_{\ell'}}(\bar p^{(i)})(\tau_{n_{\ell'}n_\ell}((K-p_i)^2)
Y_{n_\ell}(\bar q^{(j)})\, ,
\nonumber\\[2mm]
T_3^{\sf conn}&=&\sum_{i,j=1}^3
\sum_{n_{\ell'''}n_{\ell''}n_{\ell'}n_\ell}
Y_{n_{\ell'''}}(\bar p^{(i)})
\tau_{n_{\ell'''}n_{\ell''}}((K-p_i)^2){\cal M}_{n_{\ell''}n_{\ell'}}(p_i,q_j)
\nonumber\\[2mm]
&\times&\tau_{n_{\ell'}n_\ell}((K-q_j)^2)Y_{n_\ell}(\bar q^{(j)})\, .\quad\quad
\en
The vectors $\bar p^{(i)},\bar q^{(j)}$ are defined through the Lorentz boost
similar to one in Eq.~(\ref{eq:parallel}). Namely, say, $p_3^\mu$ is a four-momentum of a spectator in the final state. Define now the boost $\Lambda^\mu_\nu$,
which brings the total on-shell momentum of a pair
$P_{12}^\mu=\tilde p_1^\mu+\tilde p_2^\mu$ parallel to the vector $v^\mu$. Then,
$\bar p^{(3)\mu}=\frac{1}{2}\,\Lambda^\mu_\nu(\tilde p_1^\nu-\tilde p_2^\nu)$.
It can be also shown
that $\bar p^{(i)\mu}=\bar p^{(i)\mu}_\perp=\bar p^{(i)\mu}-v^\mu(v\bar p^{(i)})$.
The quantity $\bar q^{(j)}$ is defined similarly. Finally,
\eq
Y_{n_\ell}(\bar p^{(i)})\doteq Y_{\mu_1,\cdots,\mu_\ell}(\bar p^{(i)})
=\biggl(\frac{s}{4}-m^2\biggr)^{-\ell/2} \mathscr{Y}_{\mu_1,\cdots,\mu_\ell}(\bar p^{(i)})\, ,
\en
where the tensor $\mathscr{Y}_{\mu_1,\cdots,\mu_\ell}$ describes a particle with a 
spin $\ell$:
\eq
\mathscr{Y}&=&1\, ,
\nonumber\\[2mm]
\mathscr{Y}_\mu&=&p^\mu\, ,
\nonumber\\[2mm]
\mathscr{Y}_{\mu\nu}&=&\frac{3}{2}\,p_\mu p_\nu-\frac{1}{2}\,(g_{\mu\nu}-v_\mu v_\nu)p^2\, ,
\nonumber\\[2mm]
\cdots&&
\en
and $s=(\tilde p_1+\tilde p_2)^2$.

\subsection{Quantization condition}

In order to avoid the clutter of indices, we shall write down the quantization
condition in case of the S-wave interactions only. We start by rewriting the
Faddeev equation for the particle-dimer amplitude in a finite cubic box
of size $L$ with periodic boundary conditions, where it
takes the form:\footnote{Note that $w({\bf k})$ appears in the denominator in
  Eq.~(\ref{eq:MZL}). This happens because we carry out
  the discretization of the three-momenta in the rest frame of the box.
  To this end,
  first the Lorentz-invariant integration measure (in the infinite volume)
  $\dfrac{d^3k_\perp}{2w_v(k)}$ is rewritten as $\dfrac{d^3{\bf k}}{2w({\bf k})}$ and the discretization is performed in the latter expression.}
\eq\label{eq:MZL}
   {\cal M}_L(p,q)={\cal Z}(p,q)+\frac{1}{L^3}\sum_{\bf k}\frac{1}{2w({\bf k})}\,
   \theta(\Lambda^2+m^2-(vk)^2) {\cal Z}(p,k)\tau_L(K-k){\cal M}_L(k,q)\, .\quad\quad
   \en
   Here, $k^\mu=(w({\bf k}),{\bf k})$ and $w({\bf k})=\sqrt{m^2+{\bf k}^2}$,
   and the summation is carried out over the discrete values
   ${\bf k}=\frac{2\pi}{L}\,{\bf n},~{\bf n}\in\mathbb{Z}^3$.
   Furthermore,
   \eq\label{eq:tauLP}
   \tau_L(P)=\frac{16\pi\sqrt{s}}{p(s)\cot\delta_0(s)-\dfrac{2}{\sqrt{\pi}L\gamma}\,Z_{00}^{\bf d}(1;q_0^2)}\, ,
   \en
   where $s=P^2$, $\gamma=\left(1-\dfrac{{\bf P}^2}{P_0^2}\right)^{-1/2}$,
   ${\bf d}=\dfrac{{\bf P}L}{2\pi}$,
   $q_0^2=\dfrac{L^2}{4\pi^2}\,\left(\dfrac{s}{4}-m^2\right)$,
   and
   \eq\label{eq:zetafct}
   Z_{00}^{\bf d}(1;q_0^2)&=&\frac{1}{\sqrt{4\pi}}\sum_{{\bf r}\in P_d}\frac{1}{{\bf r}^2-q_0^2}\, ,
   \nonumber\\[2mm]
   P_d&=&\{{\bf r}=\mathbb{R}^3|r_\parallel=\gamma^{-1}\biggl(n_\parallel-\frac{1}{2}\,|{\bf d}|\biggr),\,{\bf r}_\perp={\bf n}_\perp\, ,\,{\bf n}\in\mathbb{Z}^3\}\, .
   \en
   A crucial point in the above expressions is that the two-body amplitude $\tau_L$ does not depend on $v^\mu$ even in a finite volume. In order to see this, note first that the expression $p(s)\cot\delta_0(s)$ is the same in the infinite and in finite volume and is $v^\mu$-independent. Furthermore, in the infinite volume,
   the loop is given by Eq.~(\ref{eq:Is}) and is explicitly Lorenz-invariant.
   In a finite volume, the three-momentum integral in this expression has to
   be replaced by a sum. The discretization is performed in the rest frame of a box. The result is given by the L\"uscher zeta-function, which explicitly depends on the components of the vector $P^\mu$ (i.e., is not explicitly Lorentz-invariant) but not on $v^\mu$, which does not appear at any stage. A detailed
     derivation of Eq.~\eqref{eq:tauLP} along these lines can be found in appendix \ref{app:FV2body}.

   Next,
   the above expression is written down in the assumption
   that $s>0$.
   In case of $s<0$, the $\tau_L(P)$ is replaced by $\tau(P^2)$
   -- as one knows, these two quantities below the two-particle threshold differ only by the exponentially suppressed terms.
   The latter is a perfectly well-defined Lorentz-invariant
   quantity and can be written down in terms of invariant kinematic variables, without performing any boost.

The quantization condition has the form $\det A=0$, where
   \eq
   A_{{\bf p}{\bf q}}=L^32w({\bf p})\delta_{{\bf p}{\bf q}}\tau_L^{-1}(K-p)-{\cal Z}(p,q)\, ,
   \en
   and the momenta $p,q$ obey the condition
   $\Lambda^2+m^2-(vp)^2\geq 0$,  $\Lambda^2+m^2-(vq)^2\geq 0$. 
   The zeros of the determinant determine the finite-volume spectrum in an arbitrary reference frame.

   As it is well known, the symmetries of a cubic box allow one to partially
   diagonalize the quantization condition. Below, we mainly follow the
   procedure described in~\cite{Doring:2018xxx} and generalize it to the case
   of the moving frame. In the CM frame, the rotational symmetry is reduced to
   the octahedral group $O_h$, containing 48 elements. In case of the moving
   frame, ${\bf K}\neq 0$, the symmetry is further reduced to different
   subgroups (little groups) of $O_h$, each element $g$ of which leaves
   the vector
   ${\bf d}=\dfrac{L{\bf K}}{2\pi}$
     invariant: $g{\bf d}={\bf d}$.
   These symmetry groups
   and their irreps are described in
   Refs.~\cite{Bernard:2008ax,Gockeler:2012yj}, where the matrices of the different
   irreps are explicitly given.

   In order to carry out the diagonalization of the quantization condition into
   various irreps, one has first to introduce the notion of {\em shells}
   in the space of the discretized momenta ${\bf p}=\dfrac{2\pi}{L}\, {\bf n}$,
   ${\bf n}\in\mathbb{Z}^3$. In the CM frame, a shell is defined as a set of momenta
   that can be transformed into each other by the elements of the group $O_h$~\cite{Doring:2018xxx}.
   All the elements of a given shell have the same length ${\bf n}^2$ but not
   all vectors with the same length belong to the same shell.
      In case of a moving frame, the shells are defined by two invariants
   ${\bf n}^2$, ${\bf n}{\bf d}$ instead of one.

   Following Ref.~\cite{Doring:2018xxx}, we may project the driving term
   in the quantization condition onto various irreps $\mathit{\Gamma},\mathit{\Gamma}'$:
   \eq
      {\cal Z}_{\lambda\sigma,\rho\delta}^{\mathit{\Gamma}\mathit{\Gamma}'}(r,s)     
      &=&\sum_{g,g'\in {\cal G}}(T^{\mathit{\Gamma}}_{\sigma\lambda}(g))^*
   {\cal Z}(g{\bf p}_0(r),g'{\bf q}_0(s))T^{\mathit{\Gamma}'}_{\delta\rho}(g')
\nonumber\\[2mm]
   &=&\frac{G}{s_{\mathit{\Gamma}}}\,\delta_{\mathit{\Gamma}\mathit{\Gamma}'}
   \delta_{\sigma\delta}
 \sum_{g\in {\cal G}}(T^{\mathit{\Gamma}}_{\rho\lambda}(g))^*
   {\cal Z}(g{\bf p}_0(r),{\bf q}_0(s))
\nonumber\\[2mm]
   &\doteq&\frac{G}{s_{\mathit{\Gamma}}}\,\delta_{\mathit{\Gamma}\mathit{\Gamma}'}
   \delta_{\sigma\delta}{\cal Z}^{\mathit{\Gamma}}_{\lambda\rho}(r,s)\, .
\en
Here, $r,s$ label various shells,  ${\bf p}_0(r)$ and ${\bf q}_0(s)$ denote
the pertinent {\em reference momenta}, and $T^{\mathit{\Gamma}}_{\sigma\lambda}(g)$
are the matrices of a given irrep of a group ${\cal G}$ (which coincides
with the group
$O_h$ or one of its little groups). Furthermore, $G$ is the number of the
elements in this group, and $s_{\mathit{\Gamma}}$ is the dimension
of the irrep $\mathit{\Gamma}$.

The quantization condition can be diagonalized into various irreps. It has the form $\det A^{\mathit{\Gamma}}=0$, where
\eq
A^{\mathit{\Gamma}}_{\rho\sigma}(r,s)=\delta_{rs}2w_r\delta_{\rho\sigma}\tau_L(s)^{-1}
-\frac{\sqrt{\nu(r)\nu(s)}}{GL^3}\,{\cal Z}^{\mathit{\Gamma}}_{\sigma\rho}(r,s)\, ,
\en
where $\nu(s)$ denotes the {\em multiplicity} of the shell $s$, i.e.,
the number of the independent vectors in it, and $w_r=w({\bf p})$ with vector ${\bf p}$ belonging to the shell $r$. We have further used the fact
that the quantity $\tau_L(K-k)$ is invariant under the group ${\cal G}$ and,
hence, its projection onto an irrep $\mathit{\Gamma}$ produces Kronecker symbols
only.

\subsection{Comparison with the RFT approach}

Below, we shall briefly compare the relativistic quantization condition,
written down in the present paper, to the one known in the literature, see,
for instance, Ref.~\cite{Blanton:2019igq}. Following the original derivation
given in Refs.~\cite{Hansen:2014eka, Hansen:2015zga}, one ends up with an equation
that closely resembles Eq.~(\ref{eq:MZL}) with the quantization axis chosen
at $v^\mu=v_0^\mu$. It is easy to see that a sole manifestly non-invariant
ingredient of this equation is the one-particle exchange part contained
in ${\cal Z}$ (in Refs.~\cite{Hansen:2014eka,Hansen:2015zga,Blanton:2019igq}, this
corresponds to the three-particle propagator $G$). In order to render the formalism
Lorentz-invariant, the following approach was used. The three-particle propagator
was replaced by
\eq
&&\frac{1}{2w({\bf l})}\,\frac{1}{w({\bf p})+w({\bf q})+w({\bf l})-K^0}
\nonumber\\[2mm]
&\to&
\frac{1}{2w({\bf l})}\,\frac{1}{w({\bf p})+w({\bf q})+w({\bf l})-K^0}
+\frac{1}{2w({\bf l})}\,\frac{1}{w({\bf p})+w({\bf q})-w({\bf l})-K^0}
\nonumber\\[2mm]
&=&\frac{1}{b^2-m^2}\, ,
\en
where $b^\mu=p^\mu+q^\mu-K^\mu$ and ${\bf l}={\bf K}-{\bf p}-{\bf q}$.
It can be easily seen that the added piece is a low-energy polynomial and
it can be removed by adjusting the renormalization
prescription in the short-range three-particle interaction.

This approach is, however, problematic if applied to
any formalism in which the cutoff on loop momenta can be
raised arbitrarily high.
The problem arises because the
additional term did not emerge from a Feynman integral
and thus does not have correct analytic properties. In particular, it can be
seen that the contribution, coming from the integration region where
both momenta ${\bf p}$ and ${\bf q}$ are large (of order of $m$),
violates the unitarity in the infinite volume
{\em even in the low-energy region.} This can be easily verified
looking for the zeros of the expression $w({\bf p})+w({\bf q})-w({\bf l})-K^0$
for $K^0-3m=E\ll m$. In other words, the decoupling of the low- and
high-momentum regimes, which is intimately related to the analytic properties
of the amplitudes does not occur. In a finite volume, by the same token,
it can be straightforwardly verified that
the above modification of the three-particle propagator will result in a bunch
of spurious subthreshold energy levels which have nothing to do with the real
spectrum of a system in question.

All the above effects emerge, if the integration momentum exceeds
some critical value, of order of the particle mass itself. In all analysis
carried out within the RFT approach so far, the cutoff is kept lower than
this value and, hence, the above-mentioned
deficiency did not surface. However, this
also means that the cutoff cannot be made arbitrary large in a framework
with the modified three-particle propagator. On physical grounds,
one may consider such a purely kinematic
restriction on the cutoff rather counter-intuitive, since a
cutoff is usually
associated with the massive degrees of freedom that one intends to
shield away. Moreover, one might be concerned of the fact
that the maximal allowed value of the cutoff turns out to be of order of the
particle mass.
It is however likely that, by adapting the methodology introduced here,
the cutoff in the RFT approach could be raised arbitrarily high while
maintaining relativistic invariance, and in particular the Lorentz invariance of the three-particle amplitude $K_{\sf df,3}$.

In addition, we would like to mention that in the RFT
approach\footnote{It should be noted that the similar arguments apply,
  with minor modifications, to the FVU approach as well.}, imposing
a low cutoff can be also justified by the necessity of staying above the
cross-channel cut in the two-body amplitude, as well as avoiding the
pseudothreshold singularity in the K\"allen function (in the equal-mass
case which is considered here, both,
this singularity, as well as the beginning of the left-hand cut, are located
at $s=0$, where
$s$ is the pertinent Mandelstam variable in the two-body
system).
Analogous singularities could lead to $K_{\sf df,3}$ becoming
complex-valued if the definition were modified by allowing for a higher cutoff
function. In short, both the additional pole in the relativistic analog of the
three-particle propagator $G$ and the cross-channel cut must be
carefully considered in order to modify the cutoff function in the RFT
method. This might be a
formidable task in practice which the lower cutoff helps to avoid. The NREFT
approach, in its turn, allows one to circumvent all these problems in a
systematic fashion, since the two-body amplitudes constructed here
possess the right-hand cut only, the three-body force, encoded
in the effective couplings, is real by construction
for all values of the cutoff function, and the three-particle
propagator has only one pole.
A simple physical explanation for this is that the antiparticle degrees of
freedom, which are responsible for the additional (unwanted) singularities, are hidden
in the couplings of the non-relativistic Lagrangian, both in the
two- and three-particle sector. As one knows,
this is justified only for momenta which are
much smaller than the particle mass -- for momenta of order of the
mass both the two-body amplitude and the three-body potential are
{\em modified} as compared to the relativistic theory. However,
according to the decoupling theorem, the modification of the high-energy
behavior of the amplitudes can be fully compensated at low energies by
adjusting the renormalization prescription and thus does not lead to
observable consequences. Loosely speaking, extending NREFT to describe
amplitudes for momenta of order of the particle mass and beyond
can be considered as a kind of a {\em regularization,} which consistently
removes all singularities that emerge due to the presence of the
antiparticles, and the cutoff is present solely
to tame the ultraviolet behavior.
Since all low-energy singularities are associated with
particles only, the modified NREFT correctly reproduces the singularity
structure of the amplitudes, and unitarity in the two- and three-particle
sectors is obeyed at low energies.
Furthermore, a finite-volume counterpart of this statement is
that the quantization condition, based on the 
improved NREFT approach, neglects only exponentially suppressed
volume effects at $mL\gg 1$,
but makes no other approximations associated with the non-relativistic
system. The choice to drop exponentially suppressed volume effects
is common to all methods.

Last but not least, we would like to stress once more that the discussion
of the distant singularities of different diagrams, which is given
above, does not
address the main question --  namely, at which energies these singularities become physically
important and cannot be brushed under the carpet anymore.
This problem is common for all approaches since, as already mentioned,
in order to derive the quantization condition, one is forced to restrict amplitudes on the mass shell and suppress explicit antiparticle degrees of freedom.
At this moment, we do not have an answer to this very difficult question,
which will also depend on a particular physical system considered.
The perturbative studies might
provide a clue on this issue. This, however forms
a separate subject of investigations.

\section{Exploring relativistic invariant quantization condition in a toy model}
\label{sec:numerics}

We have used the relativistic invariant quantization condition, derived
within the NREFT approach
in the previous section, for producing synthetic data within a toy model.
The aim of this investigation is to verify that the spectrum, obtained in this
manner, indeed obeys the requirements, imposed by the Lorentz invariance. In this section, we shall always used the choice for the vector
$v^\mu$ parallel to the total
four-momentum of the three-particle system $K^\mu$.

In the toy model, we consider the lowest-order S-wave interactions only, both
in the two-particle as well as in the particle-dimer channels. This means that
we have only two LECs: the non-derivative 4-particle coupling that is parameterized by the two-body scattering length $a$ and the dimensionless non-derivative
particle-dimer coupling $H_0=H_0(\Lambda)$. The driving term in the Faddeev
equation is written down as
\eq
   {\cal Z}(p,q)=\frac{1}{2w_v(K-p-q)(w_v(p)+w_v(q)+w_v(K-p-q)-vK-i\varepsilon)}
   +\frac{H_0(\Lambda)}{\Lambda^2}\, ,
   \en
   and the two-body propagator is given by
   \eq
   \tau(s)=\frac{16\pi\sqrt{s}}{-\frac{1}{a}-8\pi\sqrt{s}J(s)-ip(s)}\, .
   \en
In the non-relativistic limit, the first equation reduces to its non-relativistic counterpart displayed in Refs.~\cite{Hammer:2017uqm,Hammer:2017kms}.
Furthermore,
the finite-volume modification of the second equation, which enters the
   quantization condition, is defined according to Eq.~(\ref{eq:tauLP}).
   In addition to $a,H_0$, there are two more parameters in the model: the mass
   of the particle $m$ and the cutoff $\Lambda$. In total, this yields
   three dimensionless parameters that describe the model completely --
   we measure all dimensionful parameters in the units of $m$ and assume
   $m=1$ in the following.

   As the first quick check of our approach, we have calculated the spectrum of the
   so-called {\em Efimov states} in the infinite volume. An (infinite) tower of
   such shallow states, condensing towards the three-particle threshold,
   emerges in the non-relativistic theory in the unitary limit $a\to\infty$.
   Since in the vicinity of the threshold the particles should carry very low
   three-momenta, this non-relativistic result should be readily reproduced
   in the relativistic framework. Moreover, it is known that the
   binding energies of the neighboring Efimov states,
   $B_n = 3m - E_n$, fulfill the relation:
	\begin{equation}
		\sqrt{B_n / B_{n+1}} = \exp \left(\pi/s_0\right) \approx 22.69\,,\quad\quad s_0=1.00624\, .
              \end{equation}
              This scaling has to be reproduced by the relativistic approach,
              providing a check for the latter.
              
        In the relativistic theory, we have fixed the remaining parameters
        in the unitary limit as $\Lambda = 10^4$ and $H_0(\Lambda) = 0$.
        The results, listed in Table~\ref{tab:efimov_spectrum}, are completely in line with our expectations and confirm that our approach possesses a correct non-relativistic limit in the infinite volume.

	\begin{table}
                \begin{center}
		\begin{tabular}{ccc}
			$n$ & $B_n$ & $\sqrt{B_n/B_{n+1}}$ \\
			\hline
			1 & \SI{3.32333e-1}{} & 21.93\\
			2 & \SI{6.90973e-4}{} & 22.70\\
			3 & \SI{1.34132e-6}{} & 22.69\\
			4 & \SI{2.60432e-9}{} & 22.69\\
			5 & \SI{5.05640e-12}{} &
		\end{tabular}
                \end{center}
		\caption{Binding energies of the five deepest states for $\Lambda = 10^4$ and $H_0(\Lambda) = 0$ in the unitary limit.}
		\label{tab:efimov_spectrum}
	\end{table}

Next, the calculations in a finite volume are carried out
where we go beyond the unitary limit.
The scattering length and the cutoff in the toy model are chosen as
$a = 5$ and $\Lambda = 3$, respectively (in the units of particle mass)\footnote{It can be seen that this cutoff is high enough. At a lower cutoff, one may observe some small numerical irregularities (cusps) in the energy spectrum which represent cutoff artifacts. These irregularities are completely absent in the figures presented in this section. Also, we have checked that the low-energy spectrum is independent of $\Lambda$ to a very good accuracy, if $H_0(\Lambda)$ is re-adjusted in the infinite volume to
  keep, e.g., the particle-dimer scattering length or the energy of the (shallow) bound state constant. }.
For this value of the scattering length, a shallow dimer with the energy
$E_{d} = 1.94725$ emerges in the infinite volume and the particle-dimer
threshold lies at $E_{1d} = 2.94725$, close to the three-particle threshold.
Furthermore, requiring the existence of a three-particle bound
state at $E_1= 2.6$ in the infinite volume fixes the value of the coupling
$H_0 = -0.1182689$. Another shallow three-particle bound state is found
in the infinite volume at $E_2 = 2.94671$, very close to the
particle-dimer-threshold. All energies are given in the rest frame.

\begin{figure}
          \begin{center}
	    \includegraphics*[width=0.495\textwidth]{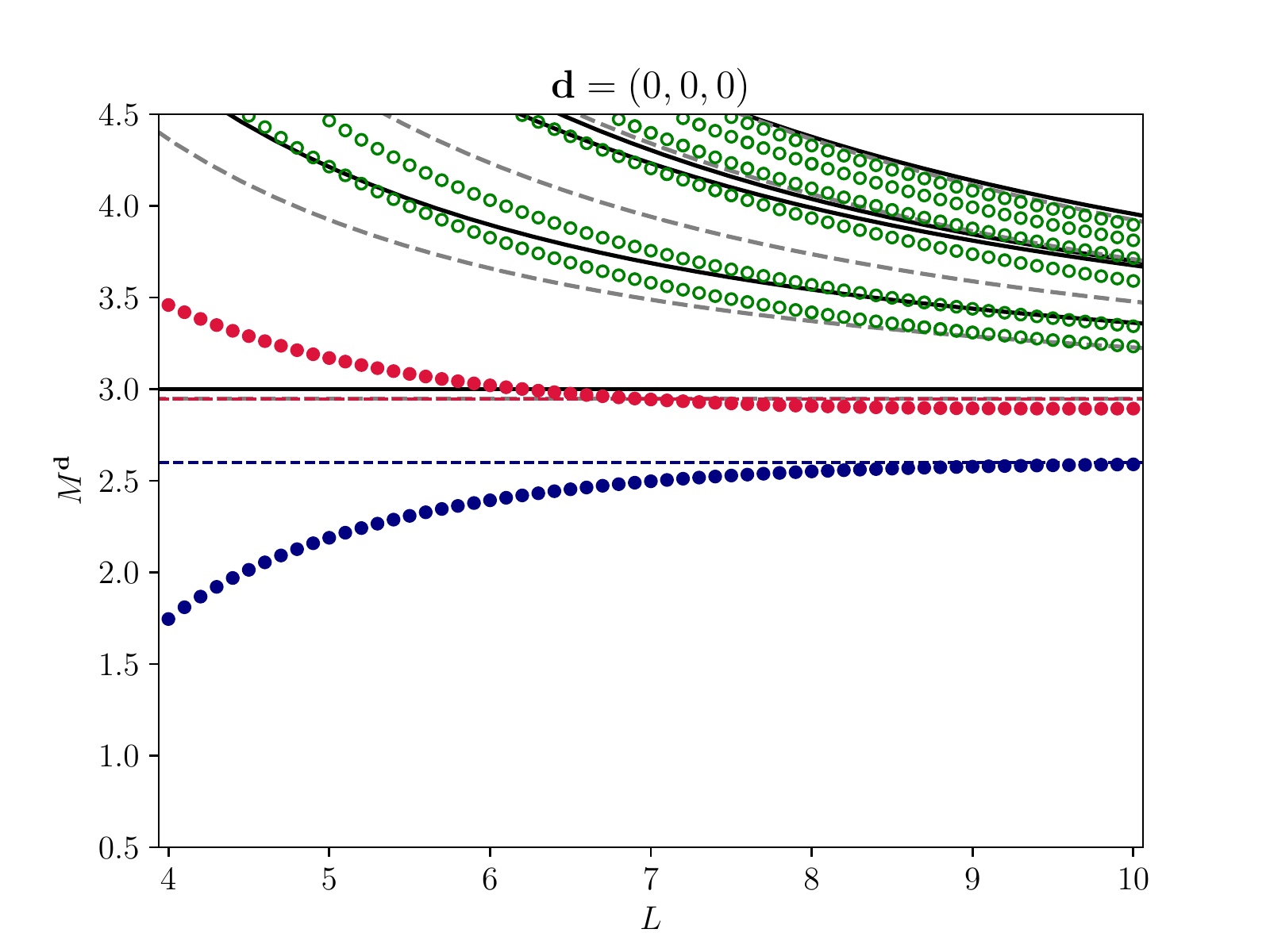}
            \includegraphics*[width=0.495\textwidth]{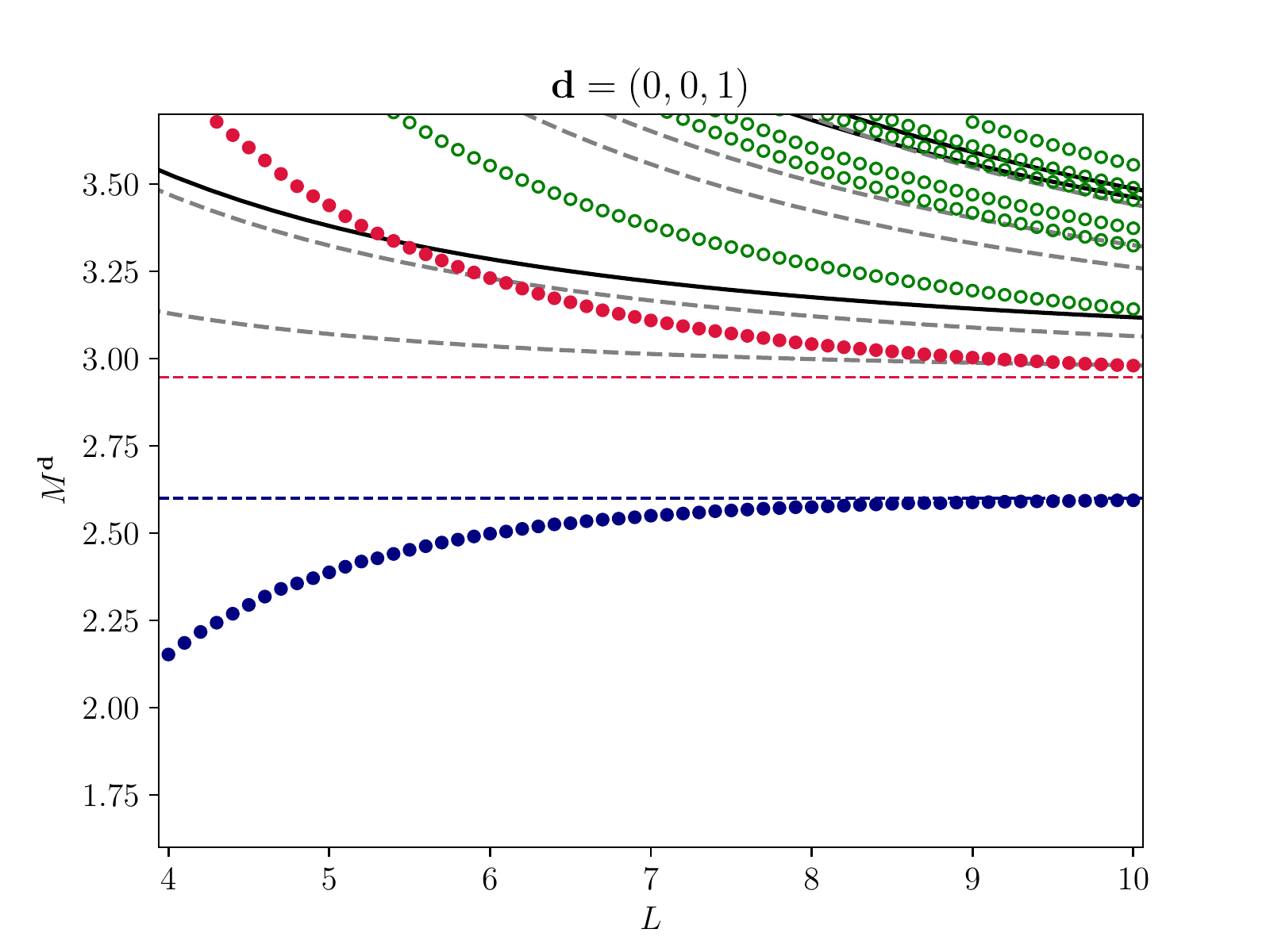}

	    \includegraphics*[width=0.495\textwidth]{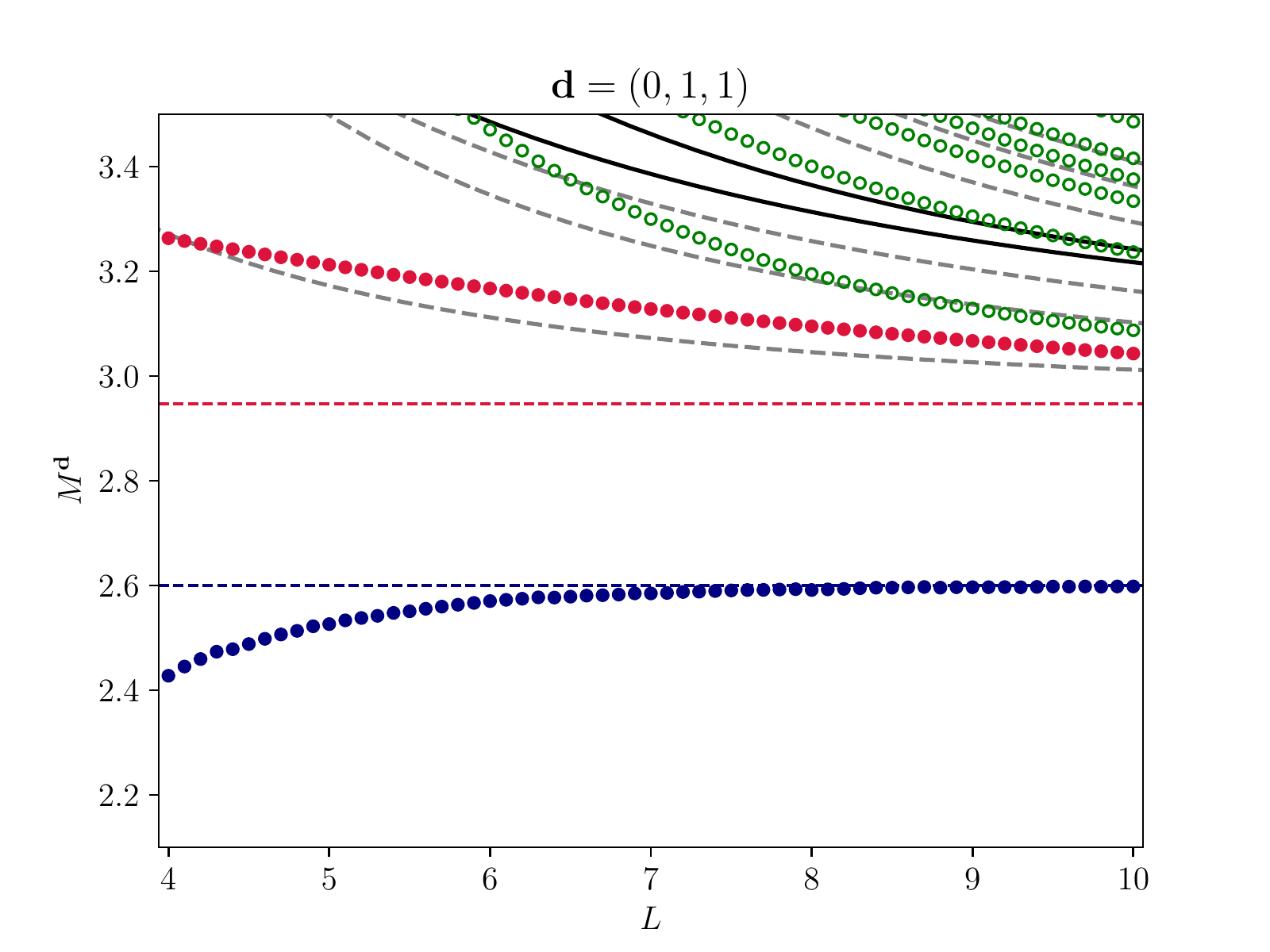}
	    \includegraphics*[width=0.495\textwidth]{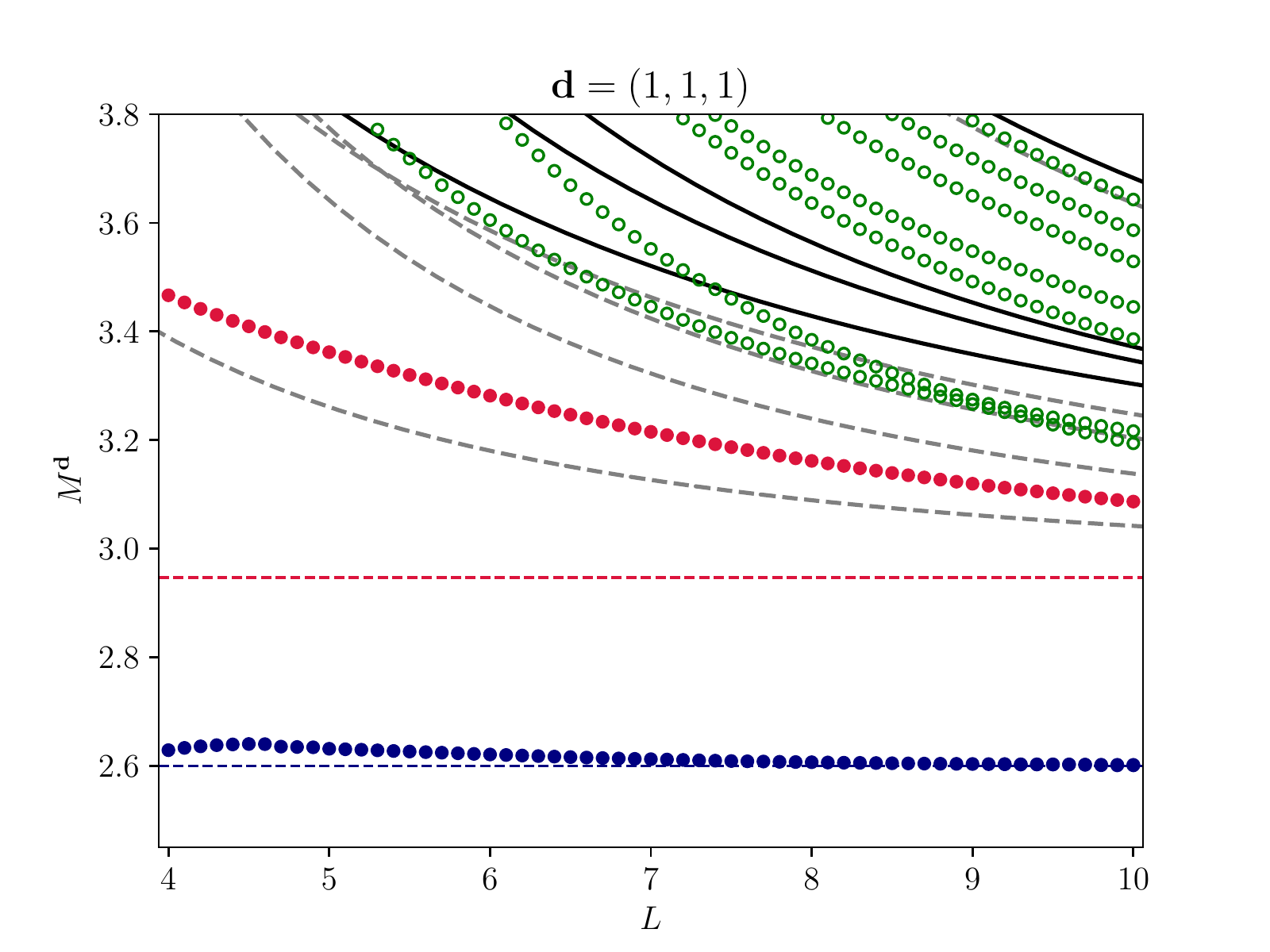}
          \end{center}
\caption{Dependence of the energy levels on the box size $L$ in the
rest frame and moving frames. Blue and red dotted curves correspond to the
energy of the deep and shallow bound states, respectively, and the green dotted
curves denote the so-called scattering states. The solid black lines and the
gray dashed lines represent the energies of three free particles and a free
particle-dimer system in a finite volume, respectively. Horizontal blue and
red dashed lines indicate the energies of the infinite-volume deep and shallow bound states. One observes an avoided level crossing (related, presumably, to the crossing of the free particle-dimer levels) in the frame
${\bf d}=(1,1,1)$ but not in the other frames. Thus, this is a purely kinematic effect.}
		\label{fig:L_dep_restframe}
	\end{figure}

Figure \ref{fig:L_dep_restframe} shows the volume dependence of the
energy spectrum in the rest frame and moving frames,
$\mathbf{d} = (0,0,1)$, $\mathbf{d} = (0,1,1)$
and $\mathbf{d} = (1,1,1)$, obtained for the above choice of the parameters,
above and below the three-particle threshold. The energy spectra are
given in terms of $M^\mathbf{d} =M^\mathbf{d}(L) = \sqrt{K_0^2 - (2\pi/L)^2 \mathbf{d}^2}$, where $K_0=K_0(L)$ are the energies in a finite volume that fulfill the
quantization condition. The lowest two levels in these figures, shown in blue and red,
correspond to the deep and shallow bound states, respectively. As seen from these figures, the shallow bound state converges to its infinite-volume limit
very slowly, as expected. Namely, for smaller $L$, the finite-volume energy
is larger than the exact infinite-volume value. With the increase of $L$ it crosses
the exact result and then approaches it from below very slowly, as $L\to\infty$.
A similar behavior was observed in the non-relativistic case, see
Ref.~\cite{Doring:2018xxx}, so the present result does not come as a surprise.
As we shall see, such an irregular behavior complicates the numerical
study of the large-$L$ limit of the shallow binding energy considerably,
especially in the moving frames where the crossing emerges at larger values
of $L$.

In order to check the relativistic invariance, we concentrate on the three-particle
bound states. Indeed, it suffices to show that the quantity $M_i^\mathbf{d}-E_i,~i=1,2$,
where the quantity $M^\mathbf{d}$ was defined above, decreases exponentially for
large values of $L$. In this case, it can be seen that the one-particle
states, obtained by solving the quantization condition, obey the relativistic
dispersion law up to the exponentially suppressed corrections. This is exactly
the result one is looking for.

\begin{figure}
\begin{center}
          \includegraphics*[width=0.495\textwidth]{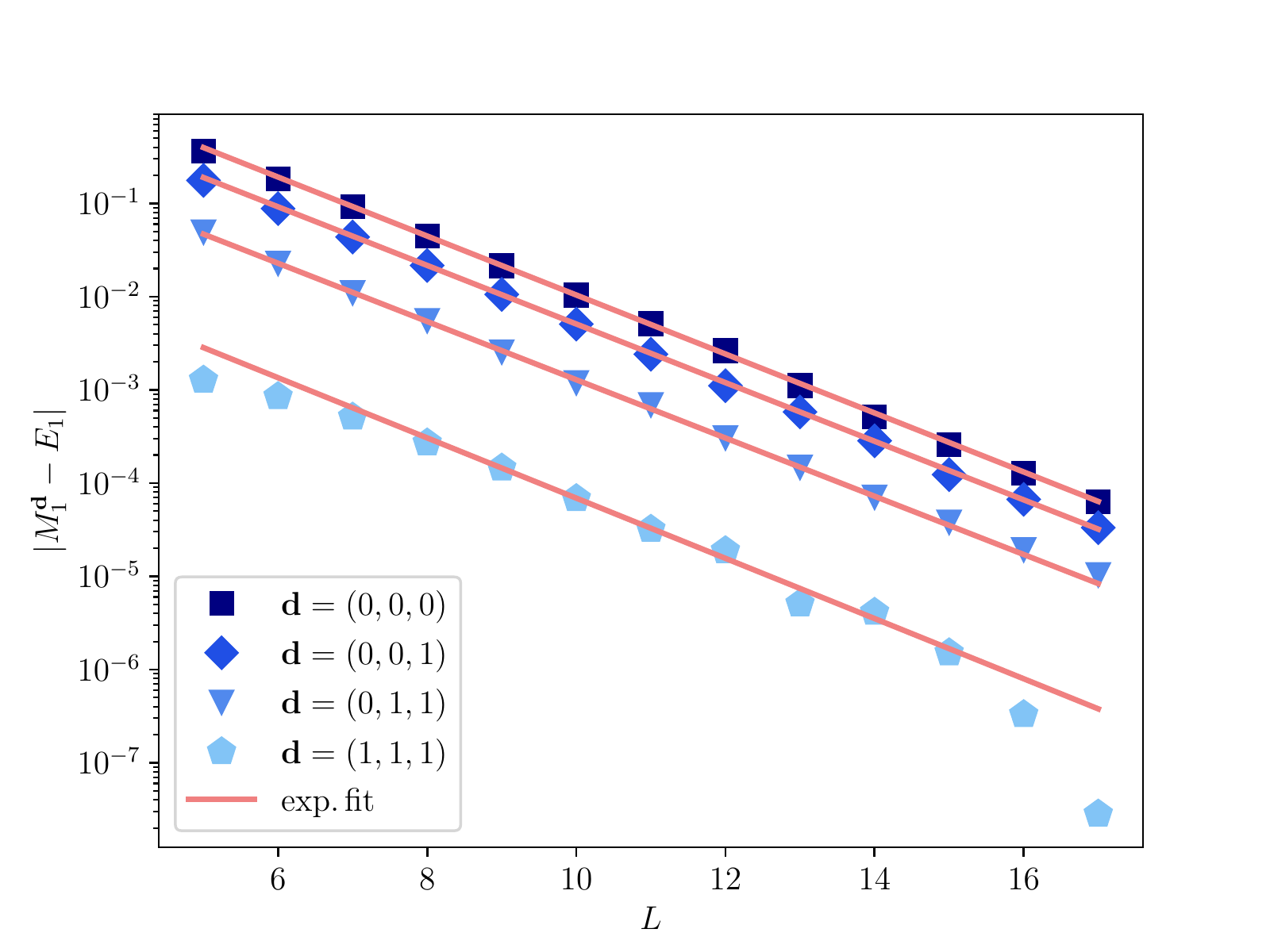}
	  \includegraphics*[width=0.495\textwidth]{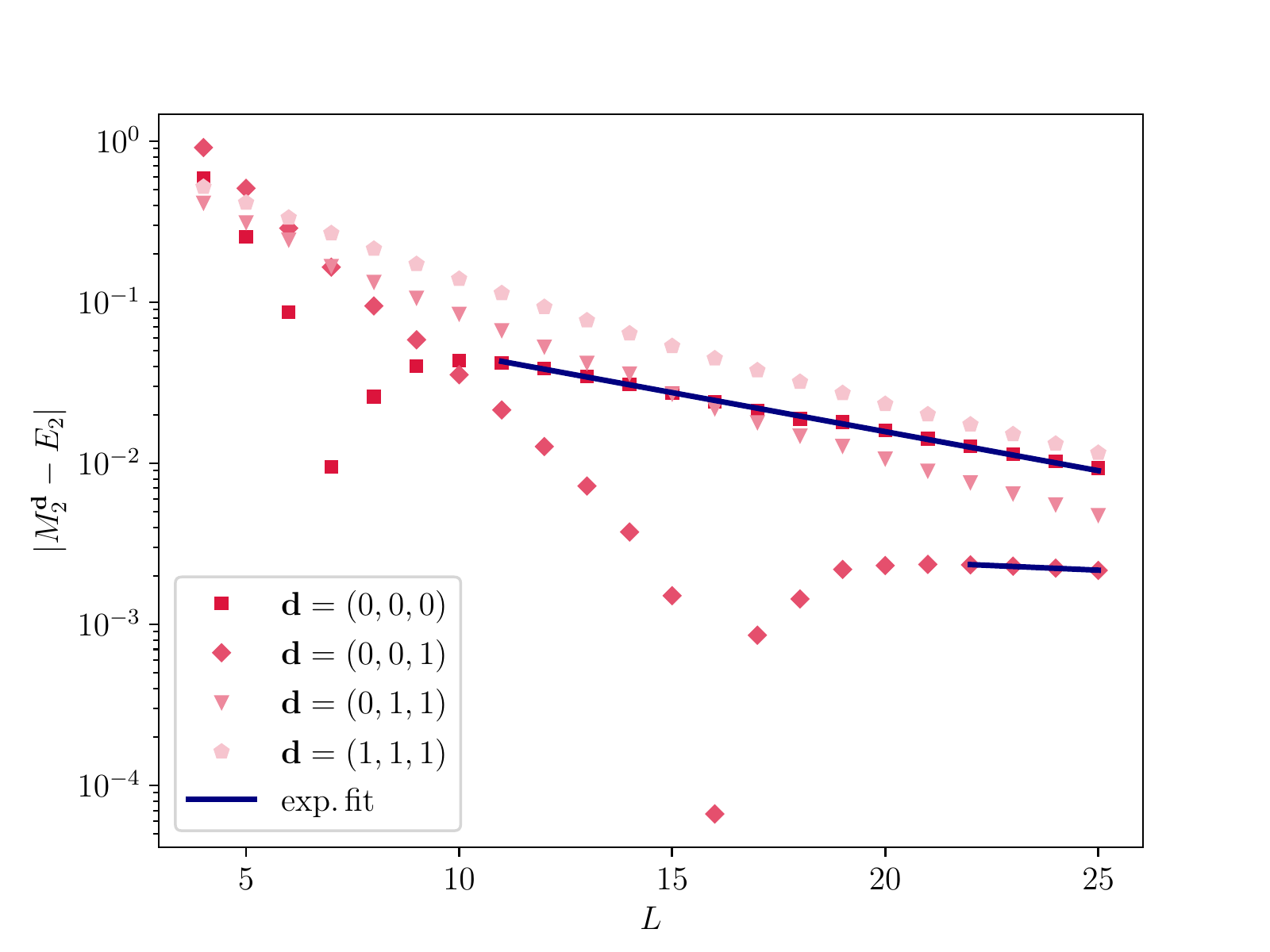}
\end{center}
\caption{The difference between the finite- and infinite-volume
  binding energies for the deep (left panel) and shallow (right panel)
  bound states. Note that for a better visibility, we have divided the energy
  shift of the deep bound state,
  corresponding to ${\bf d}=(1,1,1)$, by a factor $25$. Otherwise, the
  data for ${\bf d}=(0,1,1)$ and ${\bf d}=(1,1,1)$ would nearly overlap.}
               		\label{fig:Mdi-Ei}
                \end{figure}

\begin{figure}        
\begin{center}
          \includegraphics*[width=0.495\textwidth]{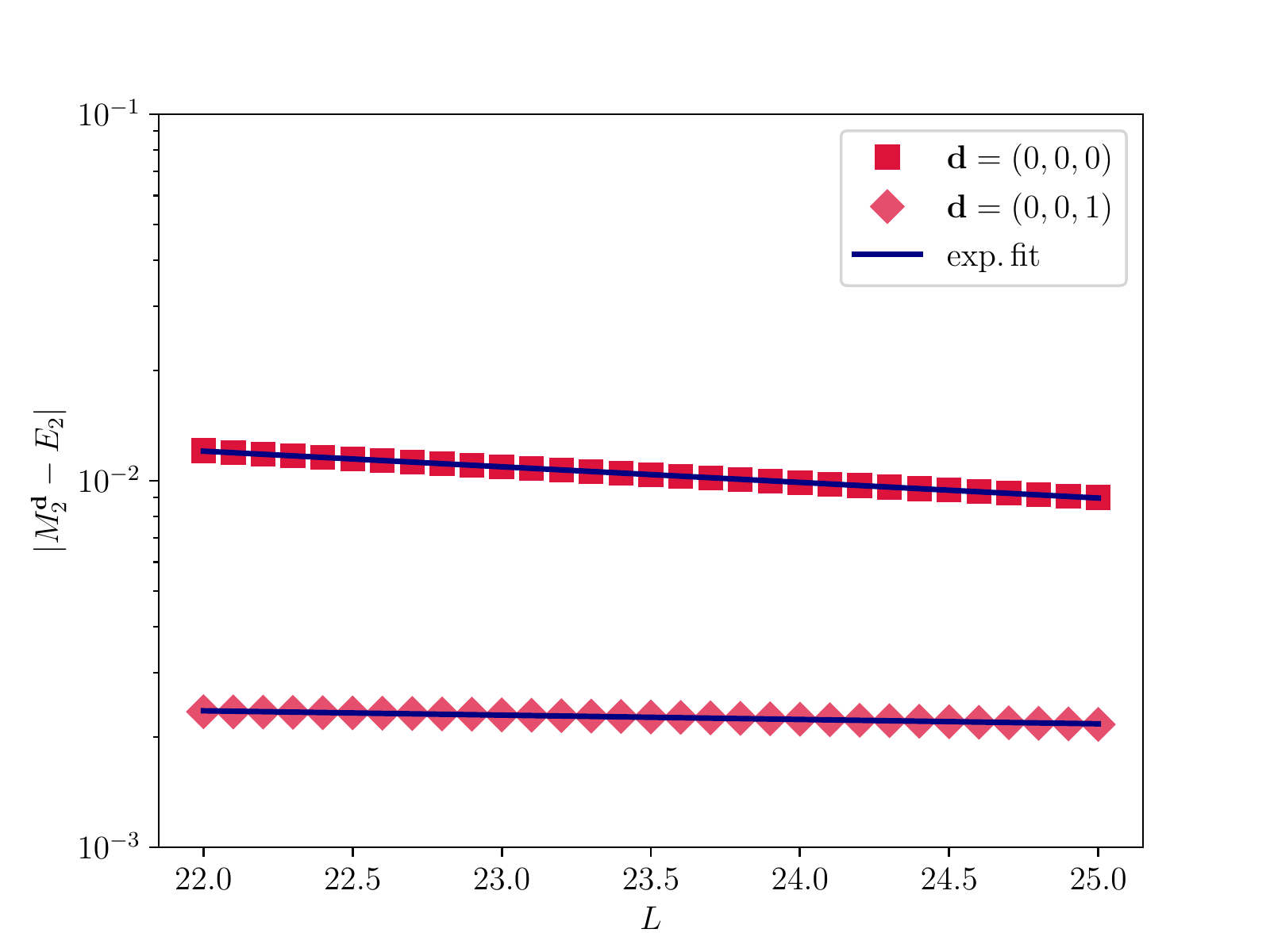}
\end{center}
\caption{The differences between the finite- and infinite-volume
  binding energies for the shallow bound state at larger values of $L$.
  The straight
  lines show the results of the exponential fit.}
               		\label{fig:restframe}
                \end{figure}

The result of the calculations is shown
in Fig.~\ref{fig:Mdi-Ei}. In case of the deep state, everything works fine.
The logarithmic plot for the difference is almost a perfect straight line
that is compatible with an exponential decrease $\sim\exp(-\kappa_{\sf deep}L)$
and $\kappa_{\sf deep}\simeq 0.7$ for all frames\footnote{The irregularities in the case ${\bf d}=(1,1,1)$ for large $L$
are caused by the fact that the cutoff is not high enough. However, since
increasing the cutoff becomes quite challenging,
we have refrained from doing this.
The exponential falloff of the corrections is anyway clearly observed
for moderate values of $L$.}. The situation with the shallow
state is different. As mentioned above, the finite- and infinite-volume
energies coincide at some $L$. This is manifested by the dips in the curves
presented on the right panel. After the dip, it takes very large values of
$L$ for the curves to stabilize and show a linear behavior. In
case of the rest frame, the curve becomes almost linear after $L\simeq 12-15$,
see Fig.~\ref{fig:restframe} (here, large values of $L$ are shown).
This behavior is consistent with the
exponential decrease $\sim\exp(-\kappa_{\sf shallow}L)$
and $\kappa_{\sf shallow}\simeq 0.11,0.03$ for the frames
$\mathbf{d} = (0,0,0),(0,0,1)$.
Note that the arguments of the exponent can be different in different frames,
because the Lorentz symmetry is broken in a finite volume. Furthermore,
the dips in the $\mathbf{d} = (0,1,1),(1,1,1)$ frames occur at much larger
values of $L$. Since carrying out calculations on such large grids is
very time-consuming, we display here the results for two reference frames only. Note also that here we did not make an attempt to {\em predict} the
values of $\kappa$ in different frames. Albeit such a theoretical prediction is possible in principle, it is not relevant in the context of the problem considered in the present paper.

To summarize, in this section it was explicitly checked that the three-particle
bound states, obtained from the solution of the quantization condition,
obey the relativistic dispersion law up to the exponentially
suppressed corrections.
Recall now that the analysis of the lattice data in the three-particle
sector proceeds in two steps. At the first step, the quantities having
a short-range
nature (like the coupling $H_0$) are extracted from data. These quantities,
like the bound-state energies, can receive only exponentially suppressed
corrections and, hence, up to such corrections, one may
use the same values of these quantities in the fit of data coming from
different moving frames. This is exactly the manifestation of the
Lorentz-invariance in a finite box. At the end, as usual, one uses an explicitly
Lorentz-invariant
infinite-volume formalism to express physical observables through the
couplings $H_0,\ldots$, extracted from data.

\section{Conclusions}
\label{sec:concl}

\begin{itemize}

\item[i)]
  In this paper, we have proposed a manifestly relativistically
  invariant formulation of the three-particle quantization
  condition within the NREFT approach. It is shown that
  the higher partial waves can be consistently included
  in the formulation. The suggested framework can be readily
  used for the global analysis of lattice spectra, measured in different moving
  frames. This was already done in case of RFT and FVU approaches.

\item[ii)]
  The method, described in this paper, is very well known for decades
  in the literature and is based on the formulation
  of the three-particle problem with an arbitrary chosen quantization axis
  defined by a unit timelike vector $v^\mu$ (see, e.g., Ref.~\cite{Kadyshevsky:1967rs}). At the end, the vector
  $v^\mu$ is fixed in terms of the external momenta that renders the
  framework manifestly invariant. The most obvious choice is to take
  that vector parallel to the total three-momentum of the system, and
  we stick to this choice. It should be also mentioned that this construction
  relies on the fact that the scattering amplitudes in the various
  two-particle subsystems, which are calculated by using the dimensional regularization and
  threshold expansion, are explicitly invariant (i.e., do not depend in $v^\mu$)
  even before fixing it in terms of the external momenta. For instance,
  this property is lost if the cutoff regularization is used for the
  two-particle subsystems as well. By the same token, a similar approach will encounter
  difficulties when applied to the four-particle problem, which features
  three-particle subsystems. This issue, however, lies beyond the scope
  of the present paper.

\item[iii)]
  The choice of the quantization axis along an arbitrary timelike vector $v^\mu$
  does not affect the analytic properties of the non-relativistic amplitudes.
  Hence, there is no violation of unitarity in this approach, and spurious
  poles do not emerge from the quantization condition.

\item[iv)]
  The proposed framework has been tested within a toy model. It has been shown that
  the three-particle bound spectrum is explicitly Lorentz-invariant, i.e.,
  the finite-volume corrections to the three-particle binding energies,
  obtained in different moving frames, are exponentially suppressed in $L$.

\item[v)]
  In our opinion, it will be rather straightforward
  to adapt the proposed method for other
  approaches used in the literature (RFT and FVU). An alternative method,
  proposed within the RFT approach, can also be used. Within this method,
  a cutoff on the three-momenta cannot be moved beyond some maximal value
  of order of a particle mass, albeit all results obtained by using the cutoffs
  less than this value are still valid.

\end{itemize}

\begin{acknowledgments}
The authors would like to thank R. Brice\~no, H.-W. Hammer, M. Hansen,
M. D\"oring, M. Mai, F. Romero-L\'opez and S. Sharpe
for interesting discussions. 
The work of F.M. and A.R. was  funded in part by
the Deutsche Forschungsgemeinschaft
(DFG, German Research Foundation) – Project-ID 196253076 – TRR 110.
A.R., in addition, thanks Volkswagenstiftung 
(grant no. 93562) and the Chinese Academy of Sciences (CAS) President's
International Fellowship Initiative (PIFI) (grant no. 2021VMB0007) for the
partial financial support.
The work of J.-Y.P. and J.-J.W. was supported
by the Fundamental Research Funds for the Central Universities, 
and by the National Key R$\&$D Program of China under
Contract No. 2020YFA0406400,
and by the Key Research Program of the Chinese Academy of
Sciences, Grant NO. XDPB15.

\end{acknowledgments}

\appendix
\section{Two-body amplitude in a finite volume}
\label{app:FV2body}
This appendix proves the $v^\mu$-independence of the finite volume two-body amplitude by providing an explicit derivation of Eq.~\eqref{eq:tauLP}. Calculating the two-particle scattering amplitude in a finite volume amounts to replacing the loop integral $I$ defined in Eq.~\eqref{eq:I-infinity} by its finite volume counterpart $I_L$.
\eq
	I_L = \frac{1}{L^3}\sum_{{\bf k}} \int\frac{dk_0}{2\pi i}\,\frac{1}{2w_v(k)(w_v(k)-vk-i\varepsilon)}\,
	\frac{1}{2w_v(P-k)(w_v(P-k)-v(P-k)-i\varepsilon)}\, .\nonumber\\
        \en
        At this stage, one uses Eq.~(\ref{eq:identity}). Adding and subtracting
        the real part of the same quantity, calculated in the infinite volume, one gets:
\eq
I_L &=& \text{Re}(I(s)) + \left[ \frac{1}{L^3}\sum_{{\bf k}} - \mathcal{P} \int\frac{d^3{\bf k}}{(2\pi)^3}\right]
\nonumber\\[2mm]
&\times&\int\frac{dk_0}{2\pi i}\,\biggl\{\frac{1}{(m^2-k^2-i\varepsilon)(m^2-(P-k)^2-i\varepsilon)}+\Delta\biggr\}\, .
\en
Here, the quantity $I(s)$ and the real part thereof are given by Eqs.~(\ref{eq:Is})-(\ref{eq:Js}). Furthermore, the quantity $\Delta$ is equal to
\eq
\Delta&=&\frac{1}{m^2-k^2-i\varepsilon}\,\frac{1}{2w_v(P-k)(w_v(P-k)+v(P-k)-i\varepsilon)}
\nonumber\\[2mm]
&+&\frac{1}{m^2-(P-k)^2-i\varepsilon}\,\frac{1}{2w_v(k)(w_v(k)+vk-i\varepsilon)}
\nonumber\\[2mm]
&+&\frac{1}{2w_v(k)(w_v(k)+vk-i\varepsilon)}\,\frac{1}{2w_v(P-k)(w_v(P-k)+v(P-k)-i\varepsilon)}\, .
\en
The energy denominators in the above expression can be expanded, according to
Eq.~(\ref{eq:expanded}). The last term turns then into a low-energy polynomial. The first
two terms contain a single low-energy pole in $k^0$ each, at $k^0=\sqrt{m^2+{\bf k}^2}$ and $k^0=P^0-\sqrt{m^2+({\bf P}-{\bf k})^2}$, respectively. Integrating over
$k^0$ leads to a low-energy polynomial again\footnote{Certain care should be taken carrying out integrations in $k^0$ over the low-energy polynomials. Strictly speaking, these integrals do not exist because of the divergence arising at $|k^0|\to\infty$. In the present papers, we consistently put all such integrals to zero that can be justified, for instance, by using
  split dimensional regularization~\cite{Leibbrandt:1996np}.}. In the infinite volume, such low-energy polynomials do not contribute to the integrals over spatial components of momenta in dimensional regularization. In a finite volume, the sum minus integral over
spatial components of momenta gives a contribution that is
exponentially suppressed in the box size $L$.
Neglecting these exponential terms, it is seen that the contribution from $\Delta$
vanishes completely, and one can finally write:
\eq
I_L=J(s) + \left[ \frac{1}{L^3}\sum_{{\bf k}} - \mathcal{P} \int\frac{d^3{\bf k}}{(2\pi)^3}\right]\int\frac{dk_0}{2\pi i}\,\frac{1}{(m^2-k^2-i\varepsilon)(m^2-(P-k)^2-i\varepsilon)}\, .
\en
The explicit $v^\mu$-dependence disappears already at this stage. The subsequent steps
are pretty standard. Evaluating the integral over $k^0$ gives:
\eq
 	I_L&=& J(s) + \left[ \frac{1}{L^3}\sum_{{\bf k}} - \mathcal{P} \int\frac{d^3k}{(2\pi)^3}\right] \frac{1}{2w({\bf k})w({\bf P-k})}\frac{w({\bf k}) +  w({\bf P-k})}{(w({\bf k}) +  w({\bf P-k}))^2-P_0^2} \nonumber\\[2mm]
 	&=& J(s) + \left[ \frac{1}{L^3}\sum_{{\bf k}} - \mathcal{P} \int\frac{d^3k}{(2\pi)^3}\right] \frac{1}{4w({\bf k})w({\bf P-k})(w({\bf k}) + w({\bf P-k})-P_0)} \nonumber\\[2mm]
 	&=& J(s) + \frac{1}{4\pi^{3/2}L\gamma\sqrt{s}}Z_{00}^{\bf d}(1;q_0^2)\, .
\en
The integrands in the first and second line differ by
$\bigl[4w({\bf k})w({\bf P-k})(w({\bf k}) + w({\bf P-k})+P_0)\bigr]^{-1}$.
Since this is a low-energy polynomial in the three-momenta, it gives rise
only to the exponentially suppressed corrections.
Finally, following Ref.~\cite{Bernard:2008ax}, the
sum minus integral in the fourth line can be expressed through
the L\"uscher zeta-function, as defined in Eq.~\eqref{eq:zetafct}.

Noting that the tree-level amplitude is the same in the infinite and
finite volume, the resulting two-body S-wave scattering amplitude
in a finite volume reads as
\eq
	\tau_L(P) = \frac{1}{(T^{\sf S-wave}_{\sf tree})^{-1}-\frac{1}{2}\,I_L(P)} = \frac{16\pi\sqrt{s}}{p(s)\cot\delta_0(s)-\dfrac{2}{\sqrt{\pi}L\gamma}\,Z_{00}^{\bf d}(1;q_0^2)}\, ,
\en
where $p(s)\cot\delta_0(s)$ is given by Eq.~\eqref{eq:pcotd}.

\end{document}